\documentclass[12pt,preprint]{aastex}

\begin{document}

\title{A wavelet-Galerkin algorithm of the E/B decomposition of
CMB polarization maps}

\author{Liang Cao\altaffilmark{1,2} and Li-Zhi Fang\altaffilmark{3}}

\altaffiltext{1}{Key Laboratory for Research in Galaxies and
Cosmology, Shanghai Astronomical Observatory, Chinese Academy of
Sciences, 80 Nandan Road, Shanghai 200030, China}
\altaffiltext{2}{Graduate University of the Chinese Academy of
Science, 19A Yuquan Road, Beijing 100049, China}
\altaffiltext{3}{Department of Physics, University of Arizona,
Tucson, AZ 85721}

\begin{abstract}

We develop an algorithm of separating the $E$ and $B$ modes of the
CMB polarization from the noisy and discretized maps of Stokes
parameter $Q$ and $U$ in a finite area. A key step of the algorithm
is to take a wavelet-Galerkin discretization of the differential
relation between the $E$, $B$ and $Q$, $U$ fields. This
discretization allows derivative operator to be represented by a
matrix, which is exactly diagonal in scale space, and narrowly
banded in spatial space. We show that the effect of boundary can be
eliminated by dropping a few DWT modes located on or nearby the
boundary. This method reveals that the derivative operators will
cause large errors in the $E$ and $B$ power spectra on small scales
if the $Q$ and $U$ maps contain Gaussian noise. It also reveals that
if the $Q$ and $U$ maps are random, these fields lead to the mixing
of the $E$ and $B$ modes. Consequently, the $B$ mode will be
contaminated if the powers of $E$ modes are much larger than that of
$B$ modes. Nevertheless, numerical tests show that the power spectra
of both $E$ and $B$ on scales larger than the finest scale by a
factor of 4 and higher can reasonably be recovered, even when the
power ratio of $E$- to $B$-modes is as large as about 10$^2$, and
the signal-to-noise ratio is equal to 10 and higher. This is because
the Galerkin discretization is free of false correlations, and keeps
the contamination under control. As wavelet variables contain
information of both spatial and scale spaces, the developed method
is also effective to recover the spatial structures of the $E$ and
$B$ mode fields.

\end{abstract}


\section{Introduction}

The scalar component of primordial perturbations of the universe can
be detected by the maps of temperature fluctuations of Cosmic
Microwave Background Radiation (CMBR), while the tensor component of
the primordial perturbations has to be probed by the maps of the
Stokes parameter $Q$ and $U$ of the linear polarization of the CMBR.
A tensor field generally contains electric-like $E$-mode and
magnetic-like $B$-modes. In the linear regime, vortical mode of primordial
perturbations do not grow during the clustering of density field, and
therefore, the perturbed field initially has to be curl-free. That is, the
primordial perturbations can only yield the $E$-mode, but not $B$-mode of
the CMBR polarization field. On the other hand, $B$-mode perturbations can
be produced by gravitational waves. Therefore, extracting the $B$-mode
information from
CMBR polarization maps is crucial to verify the existence of
gravitational wave background produced at the inflationary epoch.
Moreover, gravitational lensing of clusters and hot electron
scattering of reionization would be able to yield both $E$- and
$B$-modes. To study these problems a sharp decomposition of $E$- and
$B$-modes from $Q$ and $U$ maps is required.

If both $Q$ and $U$ maps are available over the whole sky, one can
find the whole sky maps of $E$- and $B$-modes with the spherical
harmonic decomposition, because the relation between the maps of
($Q$, $U$) and ($E$, $B$) in the space spanned by bases of spin two
harmonics is local (Kamionkowsky et al. 1997; Zaldarriaga \& Seljak
1997). However, the observed maps cannot be global; it is always limited
by the contamination of our galaxy and other foreground sources. The
relation between the maps of ($Q$, $U$) and ($E$, $B$) in physical
space contains the Laplace operator, and therefore, it is non-local. The
$E/B$ decomposition with the spatially-limited maps of $Q$ and $U$
will not be unique if we lack of information of the polarization and
its derivative on the boundary of the maps.

For noiseless samples, the problem of uniqueness would be solved by
constructing orthogonal modes with window functions to fit the
requirements of boundary conditions (Lewis et al. 2002; Bunn et al.
2003; Smith 2006; Smith \& Zaldarriaga 2007). It is, however,
similar to the domain (or windowed) Fourier analysis. The result
will not be useful to study the structures in physical space (e.g.
Chiueh \& Ma 2002).

The other challenge caused by the derivative operator is because the
$Q$ and $U$ maps are discrete. Mathematically, the derivative
operators $\partial_x$ or $\partial^2_x$ are continuous linear
operators mapping functions defined in Hilbert space, while the
observed samples $Q$ and $U$ actually are defined in space spanned
by base $v_i$, $i\in I$, which is a set of finite indices. This
difference leads to large numerical errors when the discrete maps
are noisy.

The last, but not least, problem is from the smallness of
$B$-modes. On the scale of one degree order, the power of $B$-mode
caused by gravitational waves at inflationary epoch is less than
that of $E$-modes by a factor of at least 10$^2$. As the maps of
$Q$ and $U$ are random fields, the variance of the random fields
will lead to the mixing of $E$- and $B$-modes. Consequently, $E$-
and $B$-modes would be contaminated from each other. Therefore, it
is difficult to recover the power of $B$-mode if the powers of $E$
modes are much larger than that of $B$ modes.

In this paper, we develop an algorithm of the $E/B$ decomposition
based on the discrete wavelet transform (DWT) analysis, which is a
compromise between the decompositions in physical space and scale
space. The DWT analysis of the CMBR temperature fluctuation maps has
attracted much attention in the last decade (Pando et al. 1998; Sanz
et al. 1999; Mukherjee et al. 2000). Besides these points, we
especially take the advantage of the so-called wavelet-Galerkin
discretization (e.g. Louis et al. 1997), which is to approximate
derivative operator to a matrix in space spanned by wavelet bases.
For some available wavelets, the matrixes are exactly diagonal in
scale-space, and narrowly banded in spatial space. This made the
uncertainties from boundary, noises and variances are under control.
We will study the conditions, under which the information of small
$B$-mode can approximately be extracted from noisy maps of $Q$ and
$U$.

The paper is organized as follows. Section 2 presents the method of
the $E/B$ separation in the DWT space. Section 3 tests the DWT
algorithm with samples with known spatial structures. We show that
the method effectively to suppresses the uncertainties from boundary
effect and noise. It is also effective to identify spatial
structures of $E$ and $B$ fields. Section 4 presents the tests for
samples of Gaussian random field. The effect of the variance of
Gaussian random field is analyzed, especially the problem of the
mixing of $E$- and $B$-modes. Section 5 addresses the effectiveness
of the wavelet-Galerkin discretization. Finally conclusions are
given in Section 6. The DWT representation of derivative operators
are given in the Appendix.

\section{Method}

\subsection{$E/B$ separation in DWT representation}

Let us consider polarization samples in a patch of sky, which can
be approximated as a plane described by Cartesian coordinates
$(x,y)$. In this case, the fields of $E(x,y)$ and $B(x,y)$ are
related to the maps of Stokes parameters $Q(x,y)$ and $U(x,y)$ by
(Seljak 1997)
\begin{eqnarray}
\nabla^2 E(x,y) & = &
   (\partial^2_x-\partial^2_y)Q(x,y)+2\partial_x\partial_y U(x,y)\\
\nabla^2 B(x,y) & = &
2\partial_x\partial_y Q(x,y)-(\partial^2_x-\partial^2_y)U(x,y)
\end{eqnarray}
where $\nabla^2$ is 2-D Laplace $\partial^2_x+\partial^2_y$.

We first take a wavelet-Galerkin discretization of equations (1) and
(2) to rewrite these equations in the DWT space. We can assume
that the patch is a $L\times L$ square. The size of each pixel is
$L/2^J$, $J$ being a integral, one can project the maps into the DWT
space by
\begin{equation}
\epsilon^Q_{l_1,l_2}= \int Q(x,y)\phi_{J,l_1}(x)\phi_{J,l_2}(y)dx dy
\end{equation}
\begin{equation}
\epsilon^U_{l_1,l_2}= \int U(x,y)\phi_{J,l_1}(x)\phi_{J,l_2}(y)dx dy
\end{equation}
where $\phi_{J,l}(x)$ is orthogonal scaling function on scale $J$
(e.g. Fang \& Thews 1998). It is non-zero mainly in the cell in
$x$-space from $lL/2^J$ to $(l+1)L/2^J$ . The index $l$ runs from 0 to
$2^J-1$. It spans the spatial range from 0 to $L$. The variables
$\epsilon^Q_{l_1,l_2}$ and $\epsilon^U_{l_1,l_2}$ actually are the
maps of $Q$ and $U$ on scale $J$. Since the observed maps of $Q$ and $U$
are always pixelized, the projection of eqs.(3) and (4) does not
lose information if the size $L/2^J$ is the same as that of pixels
of observed samples.

One can further take a projection on eqs.(1) and (2) as
\begin{eqnarray}
\left < \nabla^2 E(x,y), \phi_{J,l_1}(x)\phi_{J,l_2}(y )\right > & =
& \left< (\partial^2_x-\partial^2_y)Q(x,y)+2\partial_x\partial_y
U(x,y),
 \phi_{J,l_1}(x)\phi_{J,l_2}(y) \right >   \\
\left <\nabla^2 B(x,y), \phi_{J,l_1}(x)\phi_{J,l_2}(y )\right > & =
& \left < 2\partial_x\partial_y
Q(x,y)-(\partial^2_x-\partial^2_y)U(x,y),
\phi_{J,l_1}(x)\phi_{J,l_2}(y)\right >.
\end{eqnarray}
With the DWT decomposition of $E(x,y)$ and $B(x,y)$
\begin{equation}
E(x,y) =
\sum_{l_1,l_2}\epsilon^E_{l_1,l_2}\phi_{J,l_1}(x)\phi_{J,l_2}(y),
\hspace{3mm} \epsilon^E_{l_1,l_2}=\int
E(x,y)\phi_{J,l_1}(x)\phi_{J,l_2}(y)dx dy
\end{equation}
\begin{equation}
B(x,y) = \sum_{l_1,l_2} \epsilon^B_{l_1,l_2}
\phi_{J,l_1}(x)\phi_{J,l_2}(y), \hspace{3mm}
\epsilon^B_{l_1,l_2}=\int B(x,y)\phi_{J,l_1}(x)\phi_{J,l_2}(y)dx dy
\end{equation}
eqs.(5) and (6) yield matrix equations
\begin{equation}
\sum_{l'_1,l'_2} [T^{(2)}_{l_1,l'_1}\delta_{l_2,l'_2}+
T^{(2)}_{l_2,l'_2}\delta_{l_1,l'_1}]\epsilon^E_{l'_1,l'_2}=\sum_{l'_1,l'_2}
\{ [
T^{(2)}_{l_1,l'_1}\delta_{l_2,l'_2}-T^{(2)}_{l_2,l'_2}
\delta_{l_1,l'_1}]\epsilon^Q_{l'_1,l'_2}
+2T^{(1)}_{l_1,l'_1}T^{(1)}_{l_2,l_2'}\epsilon^U_{l'_1,l'_2} \}
\end{equation}
\begin{equation}
\sum_{l'_1,l'_2} [T^{(2)}_{l_1,l'_1}\delta_{l_2,l'_2}+
T^{(2)}_{l_2,l'_2}\delta_{l_1,l'_1}]\epsilon^B_{l'_1,l'_2}=\sum_{l'_1,l'_2}
\{2T^{(1)}_{l_1,l'_1}T^{(1)}_{l_2,l_2'}\epsilon^Q_{l'_1,l'_2}-[
T^{(2)}_{l_1,l'_1}\delta_{l_2,l'_2}-T^{(2)}_{l_2,l'_2}
\delta_{l_1,l'_1}]\epsilon^U_{l'_1,l'_2}
\}
\end{equation}
where the matrix $T^{(n)}_{l,l}$ is given by
\begin{equation}
T^{(n)}_{l,l'}=\int \phi_{J,l}(x)\partial^n_x\phi_{J,l'}(x)dx.
\end{equation}

Obviously we can do the projection of eqs.(5) and (6) using any bases
in 2-D space $L\times L$. However, for proper wavelet scaling
functions, the integral $\int
\phi_{J,l}(x)\partial^n_x\phi_{J',l'}(x)dx$ are zero for $J\neq J'$.
This point is important for a wavelet-Galerkin discretization (see
discussion in \S 4). In this case, all quantities of eqs.(9) and
(10) are on scale $J$, and eq.(11) gives gives
\begin{equation}
T^{(n)}_{l,l'}=\frac{1}{h^n}r^{(n)}_{l-l'}
\end{equation}
where $h=1/2^J$. $r^{n}_{l-l'}$ is non-zero only in a narrow band
$|l-l'|< M$, where $M$ is an integral, depending on wavelet. For
Daubechies 6 wavelet, the non-zero coefficients $r^{n}_{l-l'}$ are
$|l-l'|\leq 4$, or $M=4$. The values of $r^{1}_{l-l'}$ and
$r^{2}_{l-l'}$ of Daubechies 6 wavelet are listed in Table 1 of
Appendix.

Thus, eqs.(1) and (2) defined in continuous space $(x,y)$ are
reduced to eqs.(9) and (10) defined in a space spanned by orthogonal
bases $\phi_{J,l_1}(x)\phi_{J,l_2}(y)$.  The discretized Eqs.(9) and
(10) are an approximation of eqs.(1) and (2).  Equations(9) and (10) do
not contain information on scales less than $L/2^J$. However, this
discretization is reasonable in the sense that it does not introduce
false correlations, or lose information of discrete datasets $Q$ and
$U$. The derivative operator on a function defined in a space spanned by
bases $\phi_{J,l_1}(x)\phi_{J,l_2}(y)$ will yield a function in the
same space. This is required by a wavelet-Galerkin discretization
(\S 4). It ensures no signal to be produced on scales less than
$L/2^J$. The eqs.(9) and (10) give a $E/B$ decomposition from observed
maps $Q$ and $U$.

It should be pointed out that not all wavelets yield $J$-diagonal
matrices like eq.(12). For instance, the popular wavelet Daubechies 4
is not suitable for this discretization, as the matrix of
derivative operator in space spanned by Daubechies 4 scaling
functions is not $J$-diagonal. More discussion on the
wavelet-Galerkin discretization will be given in \S 4. We will first
study how to develop the algorithm of the $E/B$ decomposition with
eqs.(9) and (10).

\subsection{Maps of $\mathbb{E}_{l_1,l_2}$ and $\mathbb{B}_{l_1,l_2}$}

Equations(9) and (10) can be rewritten as follows
\begin{eqnarray}
\sum_{l_1',l_2'}\mathbb{M}_{(l_1,l_2);(l_1',l_2')}\epsilon^{E}_{l'_1,l'_2}
=\mathbb{E}_{l_1,l_2},\\
\sum_{l_1',l_2'}\mathbb{M}_{(l_1,l_2);(l_1',l_2')}\epsilon^{B}_{l'_1,l'_2}
=\mathbb{B}_{l_1,l_2},
\end{eqnarray}
where the matrix $\mathbb{M}$ is
\begin{equation}
\mathbb{M}_{(l_1,l_2);(l_1',l_2')}=T^{(2)}_{l_1,l'_1}\delta_{l_2,l'_2}+
T^{(2)}_{l_2,l'_2}\delta_{l_1,l'_1}.
\end{equation}
If we use Daubechies 6, $\mathbb{E}_{l_1,l_2}$ and
$\mathbb{B}_{l_1,l_2}$ are given by
\begin{equation}
\mathbb{E}_{l_1,l_2}  = \sum_{m_1=-4}^{4}
T^{(2)}_{m_1}\epsilon^Q_{l_1+m_1,l_2}-\sum_{m_2=-4}^{4}T^{(2)}_{m_2}
\epsilon^Q_{l_1,l_2+m_2}
+\sum_{m_1,m_2=-4}^{4}
2T^{(1)}_{m_1}T^{(1)}_{m_2}\epsilon^U_{l_1+m_1,l_2+m_2},
\end{equation}
\begin{equation}
\mathbb{B}_{l_1,l_2} = \sum_{m_1,m_2=-4}^{4}
2T^{(1)}_{m_1}T^{(1)}_{m_2}\epsilon^Q_{l_1+m_1,l_2+m_2}-\sum_{m_1=-4}^{4}
T^{(2)}_{m_1}\epsilon^U_{l_1+m_1,l_2}+\sum_{m_2=-4}^{4}
T^{(2)}_{m_2}\epsilon^U_{l_1,l_2+m_2}.
\end{equation}

The equations (13) and (14) look like the matrix equations of the
DWT variables of $E$ and $B$ fields, $\epsilon^E_{l_1,l_2}$ and
$\epsilon^B_{l_1,l_2}$. Equations(16) and (17) give, the
sources $\mathbb{E}_{l_1,l_2}$ and $\mathbb{B}_{l_1,l_2}$ on the
right hand side of eqs.(13) and (14), respectively. It seems that one can separate
$E/B$ by solving the matrix eqs.(13) and (14). However,
with the coefficients $T^{(n)}_{l,l'}$ given in Appendix, we can
show
\begin{equation}
\sum_{l'_1,l'_2}\mathbb{M}_{(l_1,l_2);(l_1',l_2')} =0.
\end{equation}
That is, the matrix $\mathbb{M}$ is singular. One cannot use a
standard linear solver to solve eqs.(13) and (14). This problem, of
course, is directly related to the non-uniqueness of the solutions
$E(x,y)$ and $B(x,y)$ given by the Poisson equations (1) and (2)
without knowledge of boundary conditions.  We will not try to solve
the singular matrix equations (13) and (14), but directly use
eqs.(16) and (17) for the $E/B$ decomposition.

\subsection{$E/B$ decomposition with $\mathbb{E}_{l_1,l_2}$ and
$\mathbb{B}_{l_1,l_2}$}

The spatial resolution of the source terms $\mathbb{E}_{l_1,l_2}$
and $\mathbb{B}_{l_1,l_2}$ is the same as maps $Q$ and $U$. It can
be used to calculate the DWT power spectrum of $\nabla^2E$ and
$\nabla^2B$ fields, and other statistics. To do these, we should
first find the wavelet function coefficient (WFC) of the maps
$\mathbb{E}_{l_1,l_2}$ and $\mathbb{B}_{l_1,l_2}$ by
\begin{equation}
\tilde{\epsilon}^{\mathbb{E}}_{\bf j,l}=\sum_{\bf l'} C_{\bf j,
l,l'} \mathbb{E}_{\bf l'}
\end{equation}
\begin{equation}
\tilde{\epsilon}^{\mathbb{B}}_{\bf j,l}=\sum_{\bf l'} C_{\bf j,
l,l'} \mathbb{B}_{\bf l'}
\end{equation}
where, for simplification, we use 2-D vector notation defined by ${\bf
j}=(j_1,j_2)$ and ${\bf l}=(l_1,l_2)$, and $l_1=0...2^{j_1}-1$,
$l_2=0...2^{j_2}-1$; $j_1$, $j_2$ can be any integral less than $J$.
Index $({\bf j,l})$ refers to the cell on scale ${\bf j}$ and at
position ${\bf l}$. The ${\bf l}\times {\bf l'}$ matrix $C_{\bf j,
l, l'}$ is given by
\begin{equation}
C_{\bf j; (l,l')}=\int
\psi_{j_1,l_1}(x)\psi_{j_2,l_2}(y)\phi_{J,l'_1}(x)\phi_{J,l'_2}(y)dxdy
\end{equation}
where $\psi_{j,l}(x)$ is 1-D wavelet function referring to cell on
scale $j$ and at position $l$. $C_{\bf j, (l, l')}$ is a banded
matrix with respect to ${\bf l, l'}$. Therefore, the relation
between $\tilde{\epsilon}^{\mathbb{E}}_{\bf j,l}$,
$\tilde{\epsilon}^{\mathbb{B}}_{\bf j,l}$ and $Q$, $U$ are spatially
quasi-local.

With the WFCs, the DWT power spectrum is given by (Fang \& Feng
2000)
\begin{equation}
P^{\mathbb{E},\mathbb{B}}_{\bf j}=
\langle(\tilde{\epsilon}^{\mathbb{E},\mathbb{B}}_{{\bf
j,l}})^2\rangle
\end{equation}
where $\langle...\rangle$ is the average over all cells ${\bf l}$.
One can directly use the DWT power spectrum to measure $E$- and
$B$-modes. $P^{\mathbb{E},\mathbb{B}}_{\bf j}$ is banded Fourier
power spectrum. For a statistically homogeneous random field, the
DWT power spectrum is related to the Fourier power spectrum
$P^{\mathbb{E},\mathbb{B}}(n_1,n_2)$ of $(\mathbb{E}$,
$\mathbb{B}$) maps by
\begin{equation}
P^{\mathbb{E},\mathbb{B}}_{\bf j} = \frac{1}{L^2} \sum_{n_1,n_2 = -
\infty}^{\infty}
 |\hat{\psi}(n_1/2^{j_1})\hat{\psi}(n_2/2^{j_2})|^2
 P^{\mathbb{E},\mathbb{B}}(n_1,n_2).
\end{equation}
Clearly, $P_{\bf j}$ is banded Fourier power spectrum with the
window function
\begin{equation}
W_{\bf j}(n_1,n_2)=\frac{1}{L^2}
|\hat{\psi}(n_1/2^{j_1})\hat{\psi}(n_2/2^{j_2})|^2.
\end{equation}
Function $\hat{\psi}(n)$ is the Fourier transform of the basic
wavelet. $P_{\bf j}$ contain all valuable quantities of second
order statistics from random samples in a finite area $L \times L$
and pixel $L/2^J$. The window function may cause spurious features
and false correlation, such as aliasing effect, in the Fourier
power spectrum. With the DWT analysis, the aliasing effects can be
effectively suppressed (Fang \& Feng 2000).

\subsection{Effect of noise on power spectrum}

The maps of $Q$ and $U$ are usually noisy and can be given by
$Q+\Delta Q$, and $U+\Delta U$. The DWT variables of noisy DWT maps
are then $\epsilon^Q_{{\bf l}}+\Delta Q_{{\bf l}}$ and
$\epsilon^U_{{\bf l}}+\Delta U_{{\bf l}}$, where $\Delta Q_{{\bf
l}}$ and $\Delta U_{{\bf l}}$ are the DWT variables of $\Delta Q$
and $\Delta U$. Assuming the noise is Gaussian and statistically
homogeneous, the DWT variables $\Delta Q_{{\bf l}}$ and $\Delta
U_{{\bf l}}$ of noise have to satisfy the following statistical
properties
\begin{equation}
\langle \Delta Q_{{\bf l}}\Delta Q_{{\bf
l'}}\rangle=\sigma^2_Q\delta_{\bf l,l'}, \hspace{3mm} \langle \Delta
U_{{\bf l}}\Delta U_{{\bf l'}}\rangle=\sigma^2_U\delta_{\bf l,l'},
\hspace{3mm} \langle \Delta Q_{{\bf l}}\Delta U_{{\bf l'}}\rangle=0
\end{equation}
where $\sigma_Q$ and $\sigma_U$ are the variance of
the noise of $Q$ and $U$ maps, respectively, and are independent of $l$.

Using $\Delta Q_{{\bf l}}$ and $\Delta U_{{\bf l}}$ to replace
$\epsilon^Q_{\bf l}$ and $\epsilon^U_{\bf l}$ in eqs.(16) and (17),
we can construct the noise maps of $\Delta \mathbb{E}_{\bf l}$ and
$\Delta \mathbb{B}_{\bf l}$. First, with eq.(25) we can show
\begin{equation}
\langle \Delta \mathbb{E}_{\bf l}\Delta \mathbb{B}_{\bf
l'}\rangle=0.
\end{equation}
That is, noise eq.(25) does not cause false correlation between
$E$ and $B$ modes. This is very helpful for the $E/B$
decomposition. Second, with eqs.(16) and (17), one can find the
variance of the noise maps $\Delta \mathbb{E}_{\bf l}$ and $\Delta
\mathbb{B}_{\bf l}$ to be
\begin{equation}
\langle (\Delta \mathbb{E}_{\bf l})^2\rangle=N^{(1)}\sigma^2_U +
N^{(2)}\sigma_Q^2
\end{equation}
\begin{equation}
\langle (\Delta \mathbb{B}_{\bf l})^2\rangle=N^{(1)}\sigma^2_Q +
N^{(2)}\sigma_U^2
\end{equation}
where
\begin{equation}
 N^{(1)}=\left
[2\sum_{m=1}^4(T^{(1)}_m )^2 \right ]^2,  \hspace{5mm}
N^{(2)}=2\sum_{m=1}^4 [ T^{(2)}_m ]^2.
\end{equation}
$N^{(1)}$ and $N^{(2)}$ are from the terms containing
$T^{(1)}_{l-l'}$ and $T^{(2)}_{l-l'}$, respectively,  in eqs.(16) and (17). For
Daubechies 6 wavelet, $\sqrt{N^{(1)}}\simeq 1.2$ and
$\sqrt{N^{(2)}}\simeq 5$. That is, in the Daubechies 6 DWT algorithm
[eqs(16) and (17)], the operator of derivative $\partial$ does not
significantly change the level of the noise, while the operator for
$\partial^2$  leads to an increase of the variance by a factor of 5
with respect to the variance of $\Delta Q$ and $\Delta U$ maps. This
shows that derivative will generally amplify the effect of noise.
However, the matrix $T^{(n)}_{l,l'}$ is exactly diagonal with
respect to $j$, the derivative operator in the DWT representation
does not transfer the noise from one scale to others. In this sense,
we have a handle on the noise.

As noise and signal are statistically uncorrelated, the power
spectrum of $\mathbb{E}_{\bf l}$ and $\mathbb{B}_{\bf l}$ can be
reconstructed by subtracting the power of noise as
\begin{equation}
P^{E}_{\bf j} = P^{\mathbb{E}}_{\bf j }- P^{\Delta \mathbb{E}}_{\bf j}
\end{equation}
\begin{equation}
P^{B}_{\bf j} = P^{\mathbb{B}}_{\bf j }- P^{\Delta \mathbb{B}}_{\bf j}
\end{equation}
where $P^{\mathbb{E}}_{\bf j }$ and $P^{\mathbb{B}}_{\bf j }$ are
the DWT power spectrum of maps $\mathbb{E}_{\bf l}$ and
$\mathbb{B}_{\bf l}$ given by eqs.(16) and (17), respectively,
using noisy $Q$ and $U$. $P^{\Delta \mathbb{E} }_{\bf j}$ and
$P^{\Delta \mathbb{B} }_{\bf j}$ are the DWT power spectra of noise
maps $\Delta \mathbb{E}_{\bf l}$ and $\Delta \mathbb{B}_{\bf l}$.
The algorithm of subtracting the noise DWT power spectrum $P^{\Delta
\mathbb{E}}_{\bf j}$ and $P^{\Delta \mathbb{B}}_{\bf j}$
scale-by-scale is similar to the subtraction of shot noise power
from the DWT power spectrum of galaxy survey (Fang \& Feng 2000).

\subsection{The effect of boundary}

For a sample of finite area, the DWT power spectrum analysis does
not need a window function to treat the spatial domain. The effect
of boundary can effectively be reduced by dropping the DWT
variables related to cells $(j,l)$ located on or near the boundary
(Pando \& Fang 1998). When derivative operators, $\partial_x$,
$\partial_y$, are involved, the boundary effect would be more
serious, because the matrix of derivative operators in the DWT
representation is not exactly diagonal with respect to the spatial
index $l$. Nevertheless, the matrix $T^{(n)}_{l,l'}$ [eq.(11)] is
narrowly banded, the effect of boundary can still be reduced by
dropping boundary modes.

\section{Tests with samples having known spatial structures}

To test the DWT algorithm developed in \S 3, we consider, in this
section, samples with given spatial structures, and compare the maps
$\mathbb{E}_{l,l'}$ and $\mathbb{B}_{l,l'}$ given by eqs.(16) and
(17) with that directly calculated from $E$ and $B$. This comparison
is only to test the discretization of derivative operator, but says
nothing about the amount of information loss associated with the
algorithm.

\subsection{Samples}

We use two scalar functions $\psi_E(x,y)$ and $\psi_B(x,y)$ to produce $E$
and $B$ maps in 2-D space by the following way
\begin{equation}
E=-\nabla^2 \psi_E, \hspace{5mm} B=-\nabla^2 \psi_B.
\end{equation}
One can then produce the DWT variables $\epsilon^E_{l,l'}$ and
$\epsilon^B_{l,l'}$ with eqs.(7) and (8). With these results, we can
further produce the maps of $\mathbb{E}_{l,l'}$ and
$\mathbb{B}_{l,l'}$ with eqs.(13) and (14). Thus, for given scalar
functions $\psi_E(x,y)$ and $\psi_B(x,y)$, we have the samples
$\mathbb{E}_{l,l'}$ and $\mathbb{B}_{l,l'}$, and then, the the DWT
power spectrum and other statistical properties of maps
$\mathbb{E}_{l,l'}$ and $\mathbb{B}_{l,l'}$. The ratio between the
powers of $E$- and $B$-modes can be adjusted by the ratio between
the functions $\psi_E$ and $\psi_B$.

On the other hand, using the function $\psi_E(x,y)$ and
$\psi_B(x,y)$, one can produce the samples of the Stokes parameters
$Q(x,y)$ and $U(x,y)$ maps by
\begin{eqnarray}
Q(x,y) & = & (\partial_x\partial_x-\partial_y\partial_y)\psi_E(x,y)-
     2\partial_x\partial_y\psi_B(x,y), \\
U(x,y) & = & 2\partial_x\partial_y\psi_E(x,y)+
    (\partial_x\partial_x-\partial_y\partial_y)\psi_B(x,y).
\end{eqnarray}
We add Gaussian white noise in the $Q$ and $U$ maps pixel-by-pixel
with signal-to-noise ratio equal to 10, 50, and 100. These $Q$ and
$U$ maps are used as the simulation of observed samples.

With noisy maps $Q$ and $U$, we can produce the variables
$\epsilon^Q_{l,l'}$ and $\epsilon^U_{l,l'}$ by the projection of
eqs.(3) and (4). Finally, we have maps $\mathbb{E}_{l,l'}$ and
$\mathbb{B}_{l,l'}$ using eqs.(16) and (17). Thus, we can test the
algorithm by comparing the statistics of the maps
$\mathbb{E}_{l,l'}$  and $\mathbb{B}_{l,l'}$ given by $Q$ and $U$
[eqs.(33) and (34)] with that directly calculated from $E$ and $B$
of eq.(32).

\subsection{Recovery of spatial structures}

The scalar functions $\psi_{E}$ and $\psi_{B}$ are taken to be sample A.)
Gaussian function $\psi_{E,B}(x,y)=a_{E,B}\exp-(x^2+y^2)/2d^2$;
sample B.) the Legendre function
$\psi_{E,B}(x,y)=a_{E,B}P_m(x)P_m(y)$. Both samples are in the area
$-0.5\leq x\leq 0.5$ and $-0.5\leq y \leq 0.5$ and pixels $512\times
512$, i.e. $J=8$. The coefficients $a_{E}$ and $a_{B}$ are used to
adjust the ratio of the powers of $\mathbb{E}_{l,l'}$ and
$\mathbb{B}_{l,l'}$. We use $a_{E}=1$ and $a_{B}=1/10$. That is, the
power of $E$-mode is larger than $B$ mode by a factor $10^2$. The
maps of $\mathbb{E}_{l,l'}$ for samples $A$ and $B$ in the central
square $32\times32$ pixels are shown Figure 1. The maps of
$\mathbb{B}_{l,l'}$ have the same shape of Figure 1, but the
intensity is weaker than Figure 1 by a factor $10$.
\begin{figure}[htb]
\begin{center}
\includegraphics[width=5.0cm]{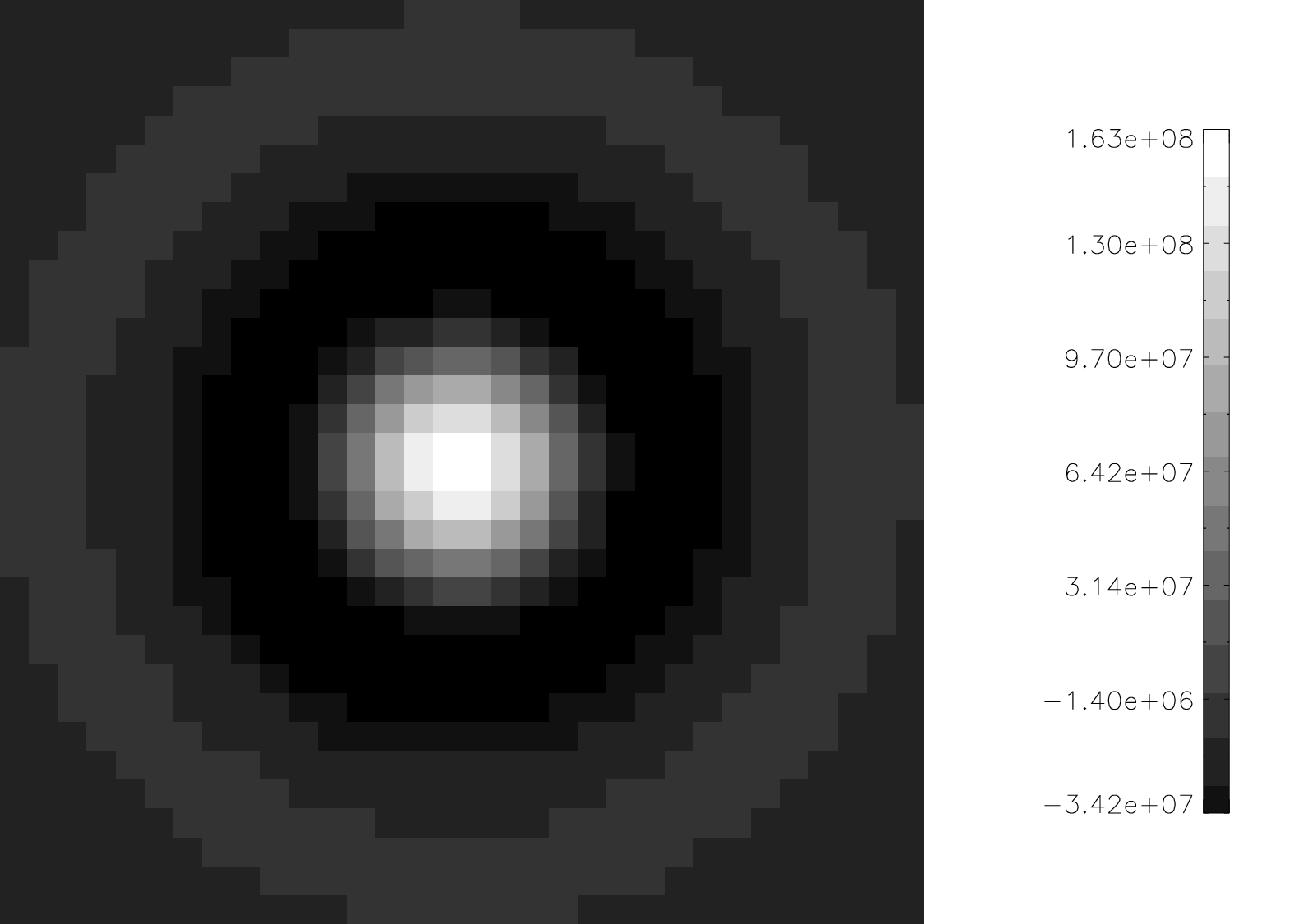}
\includegraphics[width=5.0cm]{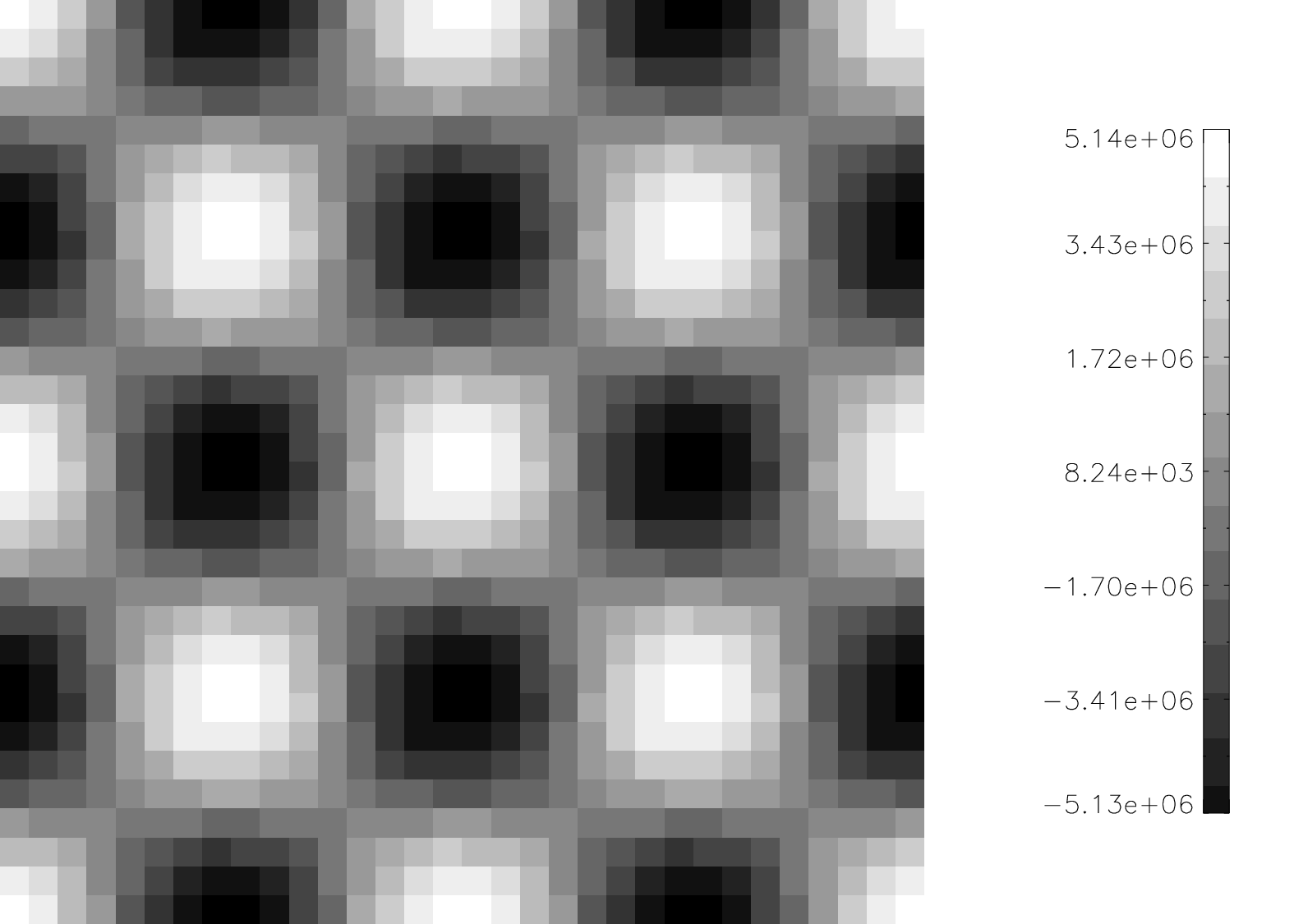}
\end{center}
\begin{center}
\vspace{-0cm} \caption{The DWT maps of $\mathbb{E}_{\bf l}$ of
sample A: $\psi_E=\exp[-(x^2+y^2)/2d^2]$, and $2d^2=1600$ (left), and
sample B: $\psi_E=\psi_B/a_B=P_m(x)P_m(y)$, and $m=100$ (right). }
\end{center}
\end{figure}
\begin{figure}[htb]
\begin{center}
\includegraphics[width=5.0cm]{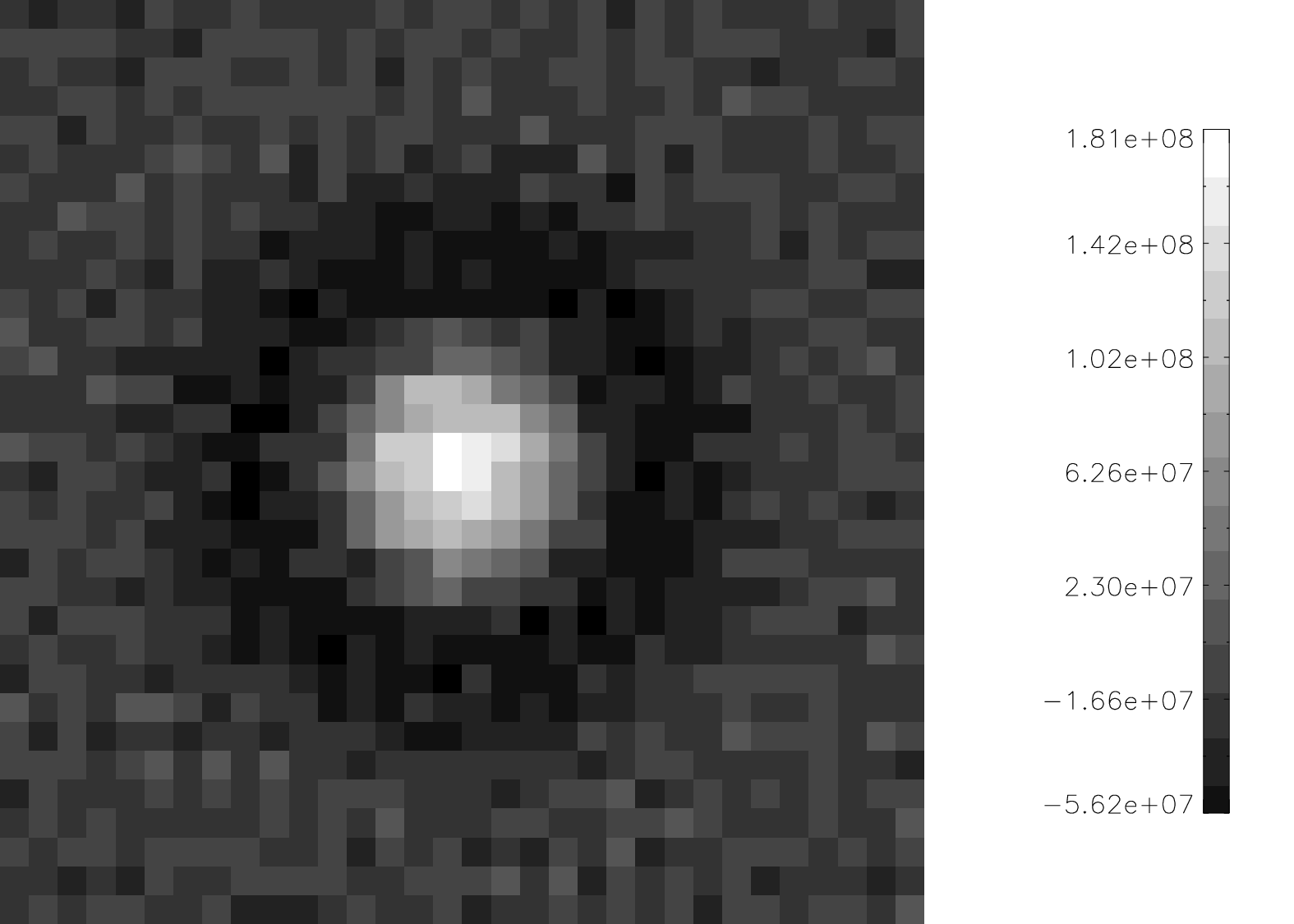}
\hspace{-0.0cm}\includegraphics[width=5.0cm]{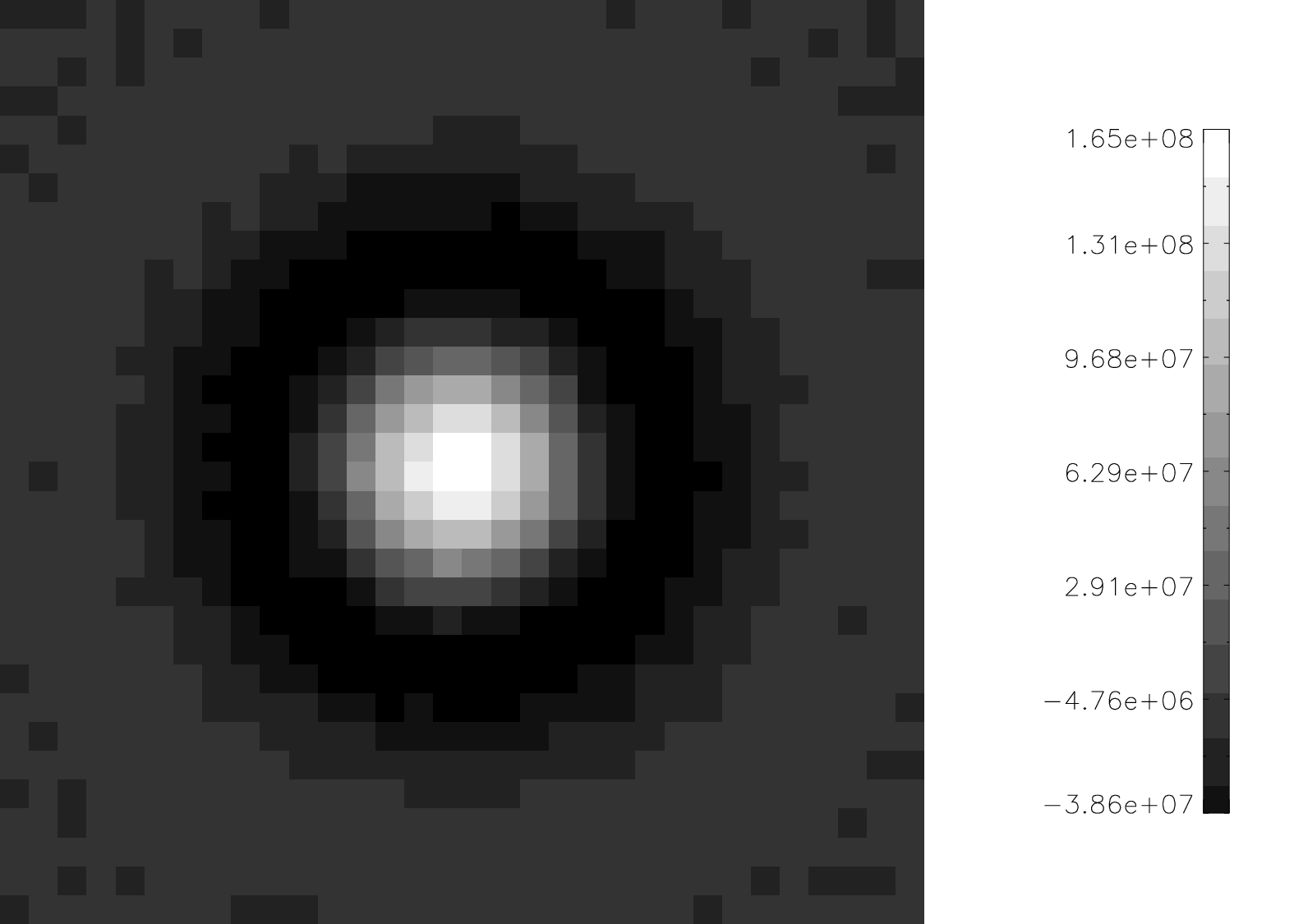}
\hspace{-0.0cm}\includegraphics[width=5.0cm]{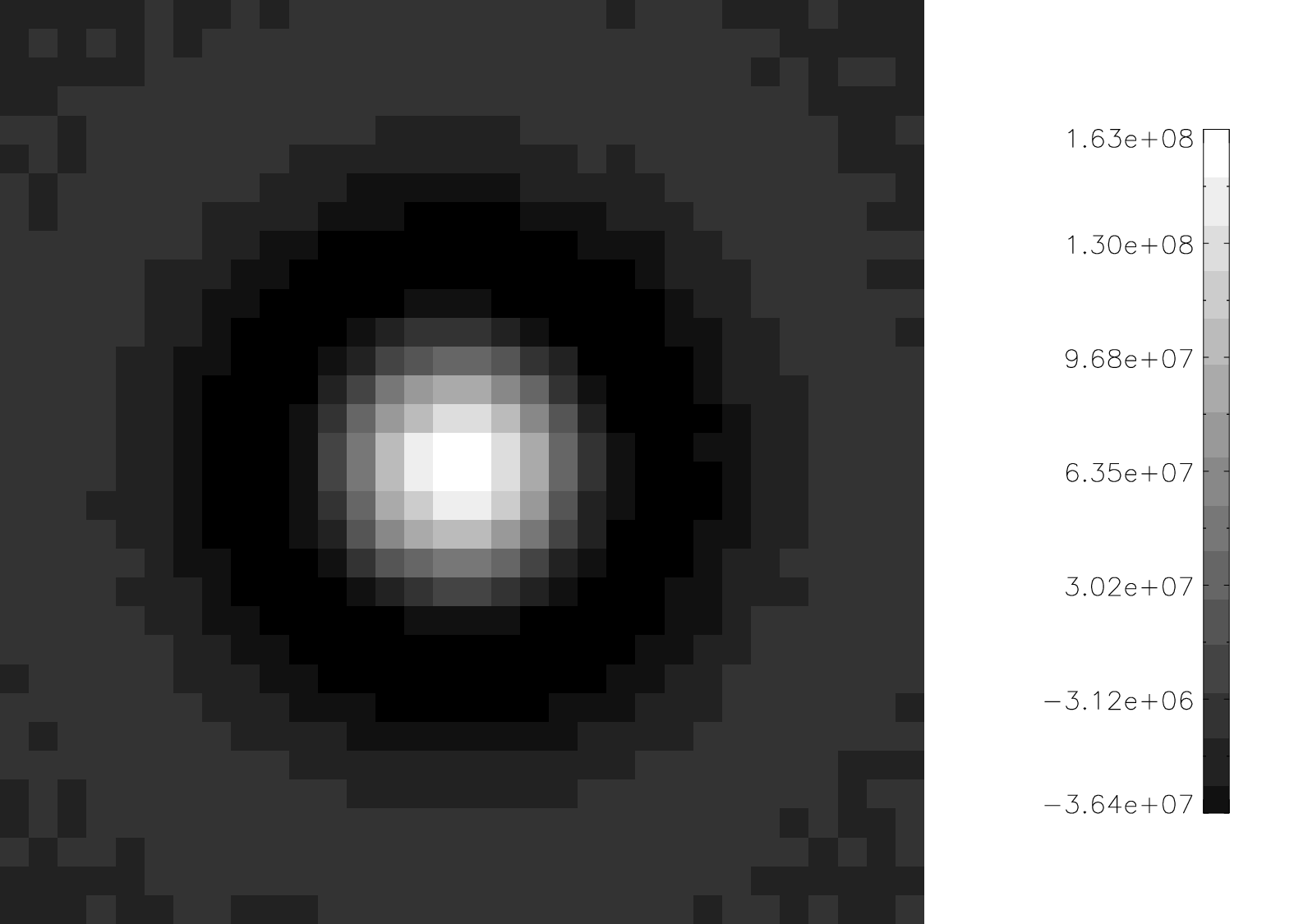}
\end{center}
\begin{center}
\includegraphics[width=5.0cm]{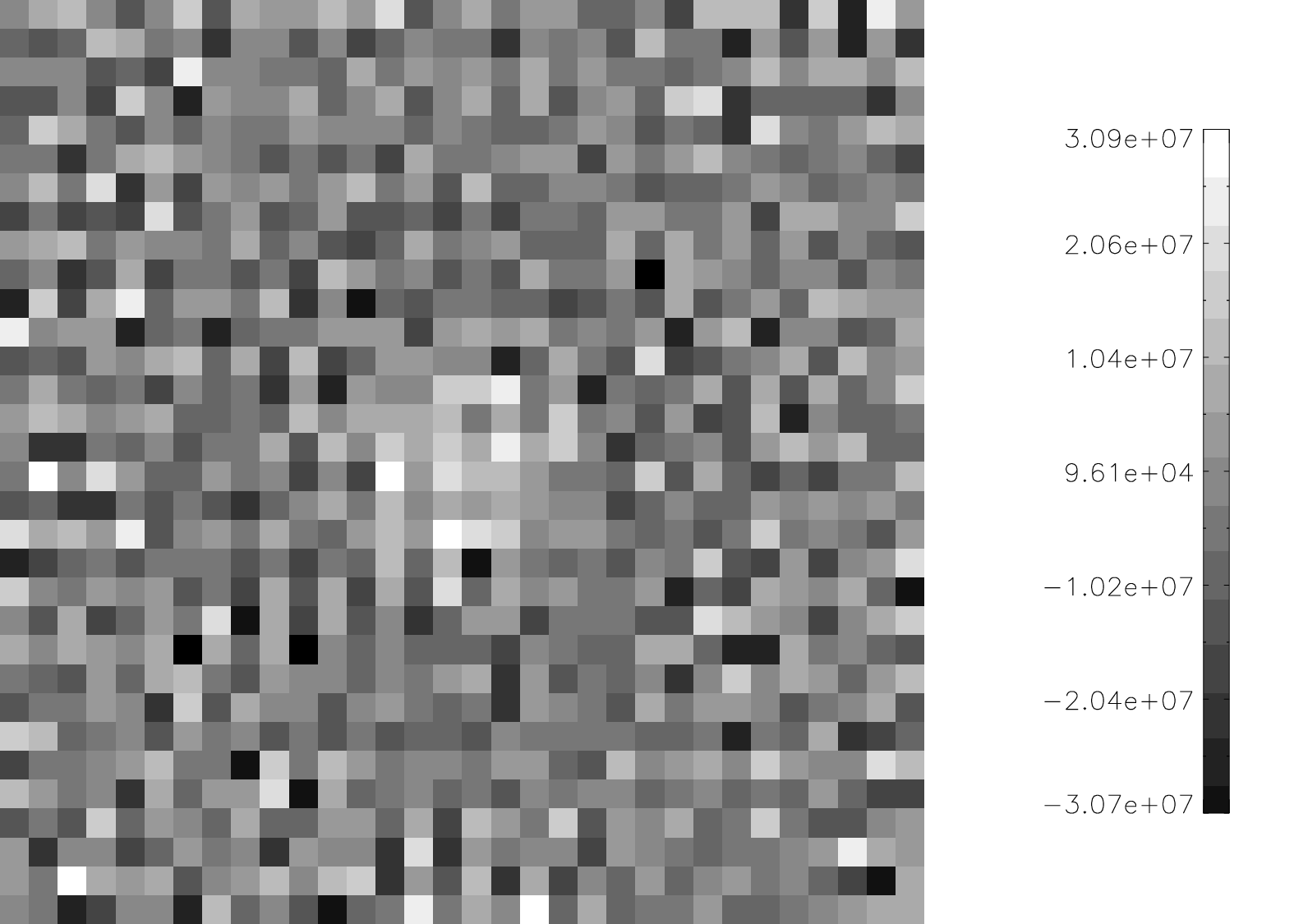}
\hspace{-0.0cm}\includegraphics[width=5.0cm]{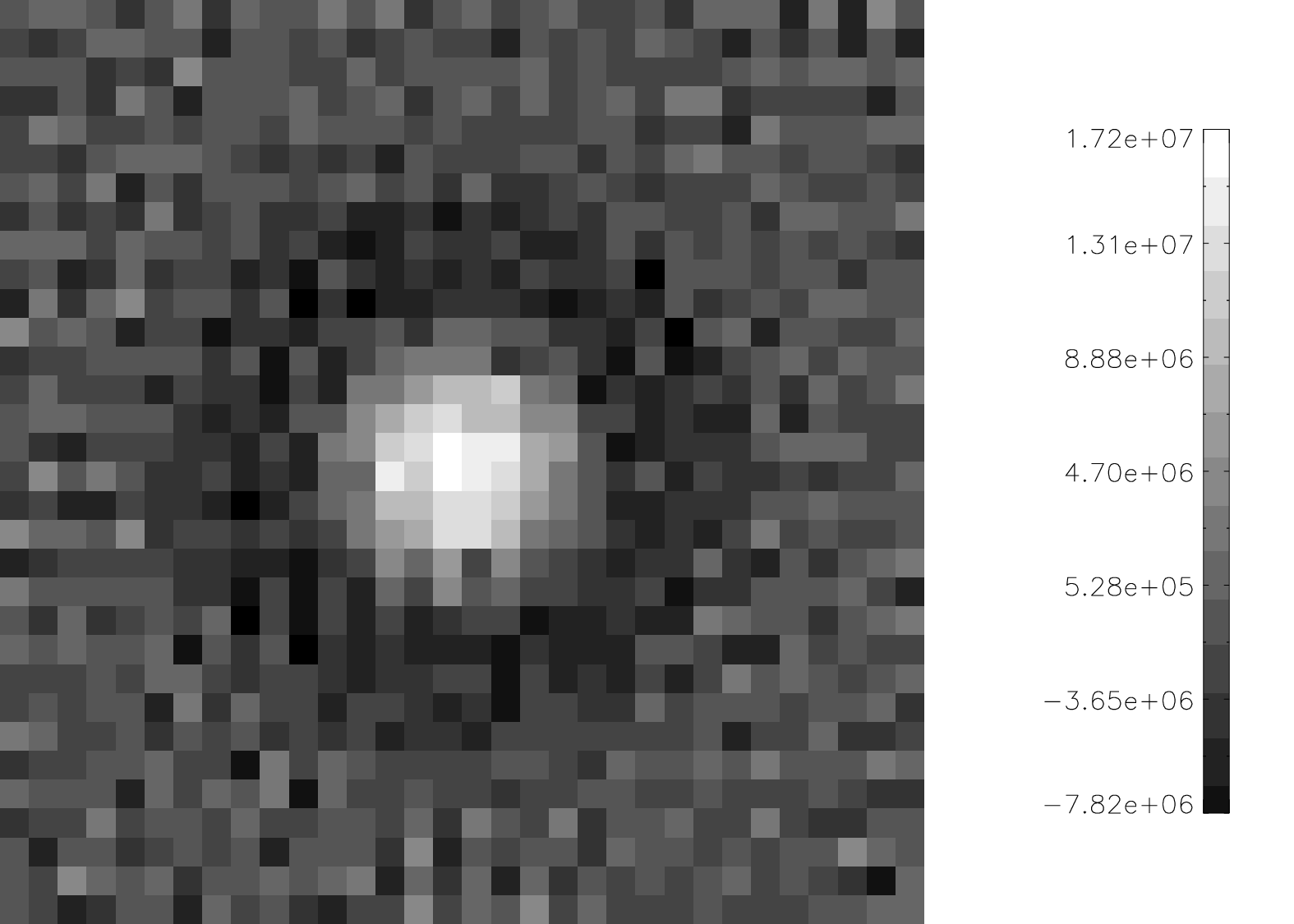}
\hspace{-0.0cm}\includegraphics[width=5.0cm]{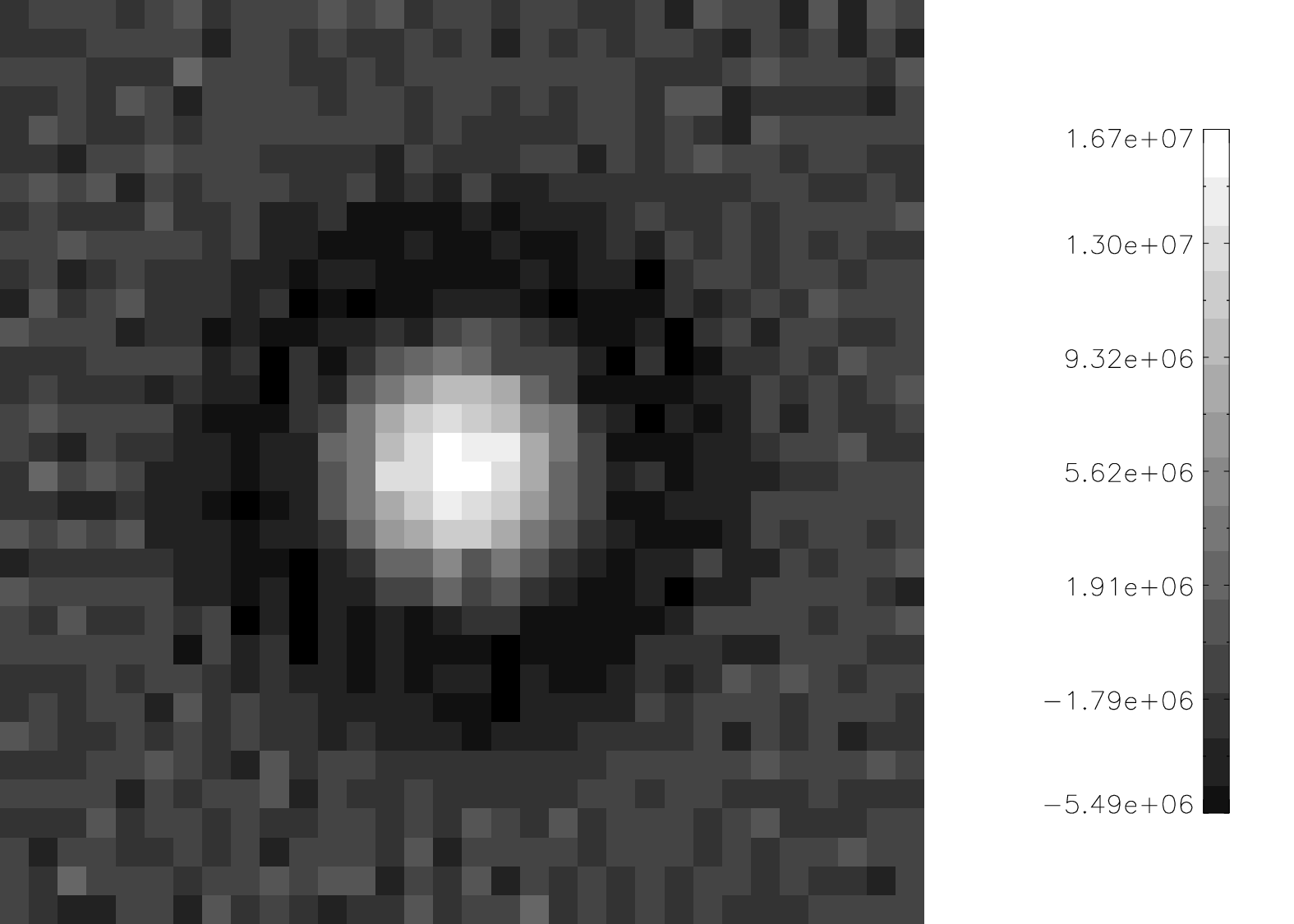}
\caption{The maps $\mathbb{E}_{l_1,l_2}$ (up panels) and
 $\mathbb{B}_{l_1,l_2}$ (bottom panel) of sample $A$, which
are recovered by eqs.(16) and (17) with noisy maps $Q$ and $U$, and
the signal-to-noise ratios are taken to be S/N=10 (left), 50
(middle) and 100 (right).}
\end{center}
\end{figure}
\begin{figure}[htb]
\begin{center}
\includegraphics[width=5.0cm]{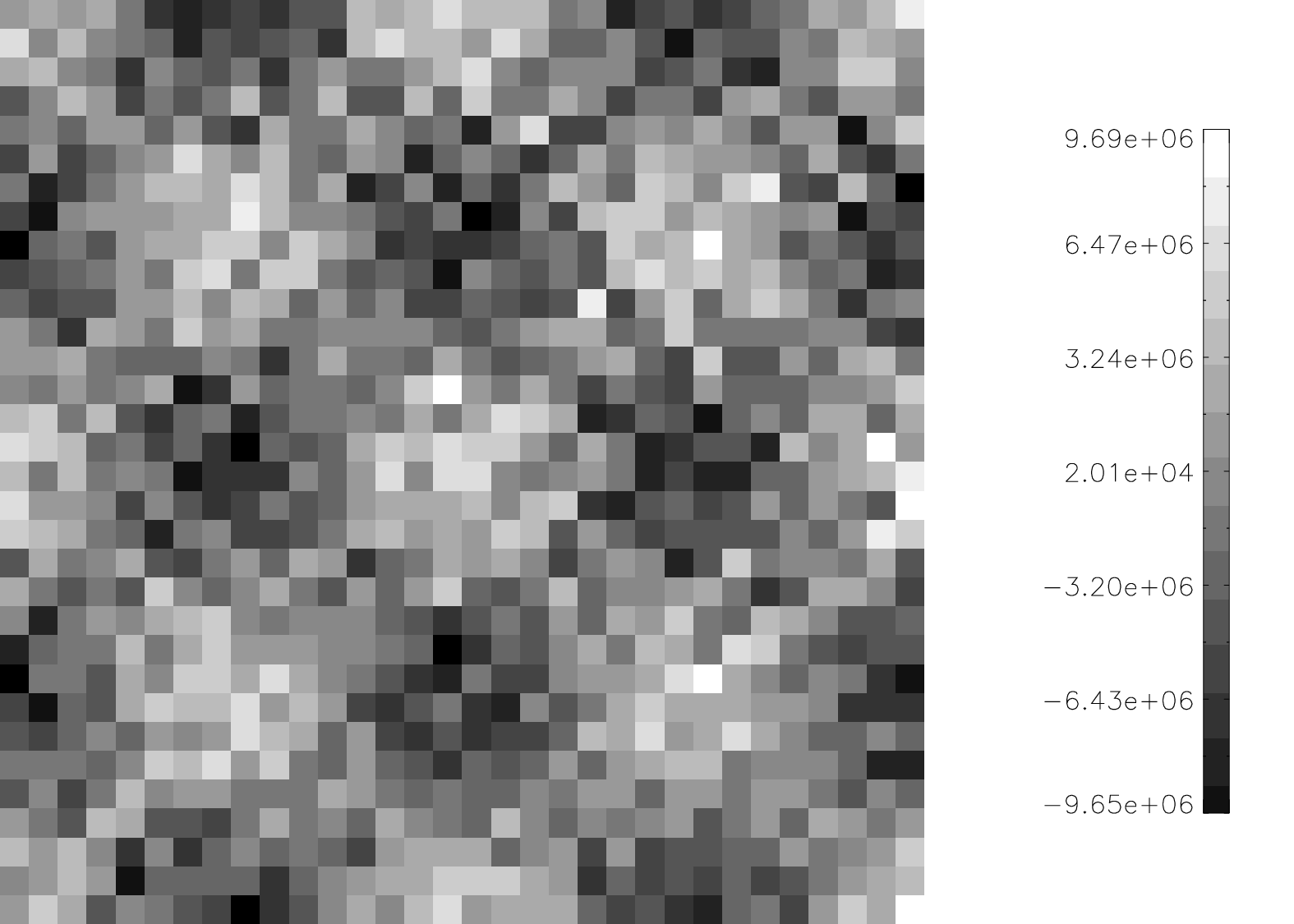}
\hspace{-0.0cm}\includegraphics[width=5.0cm]{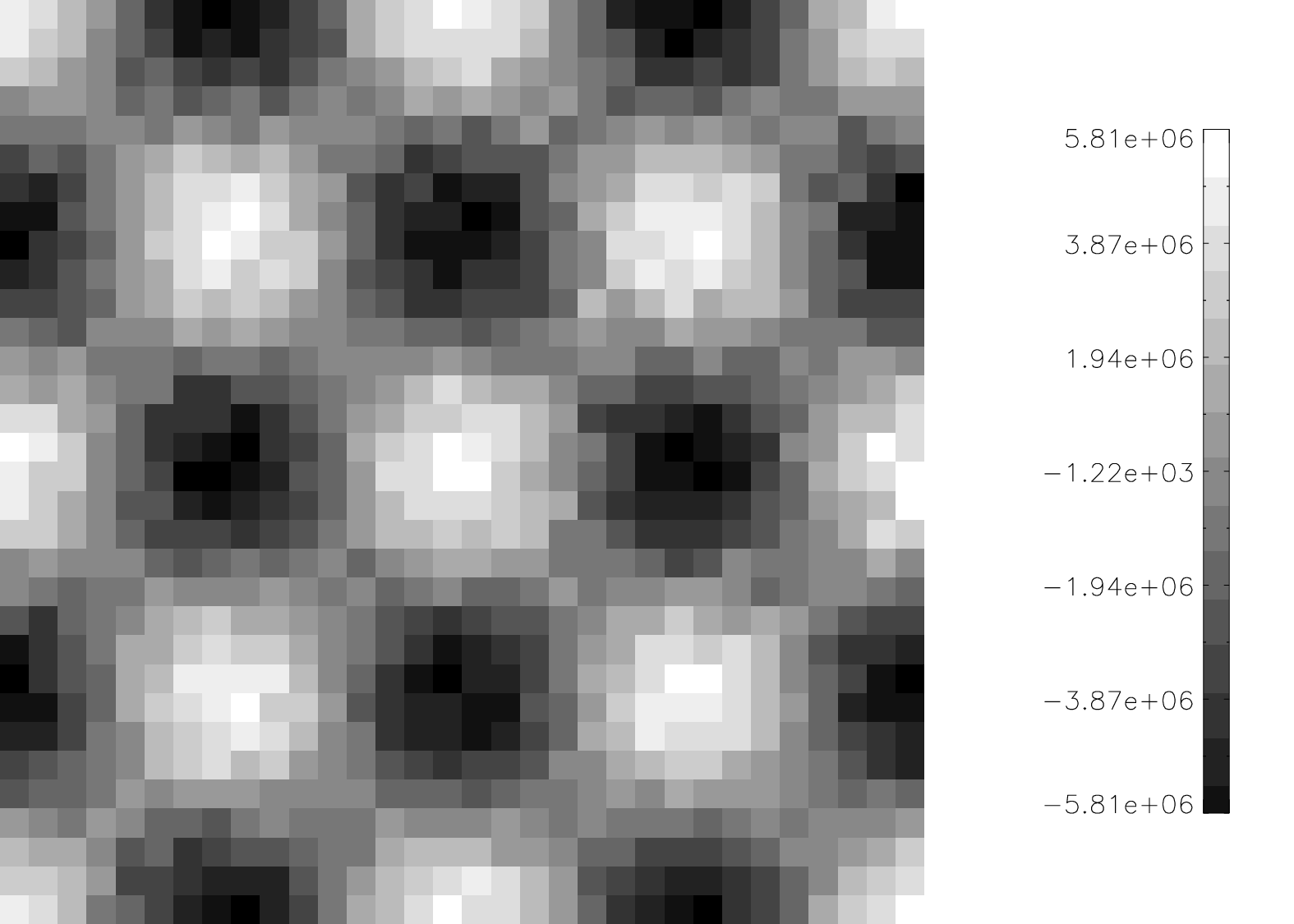}
\hspace{-0.0cm}\includegraphics[width=5.0cm]{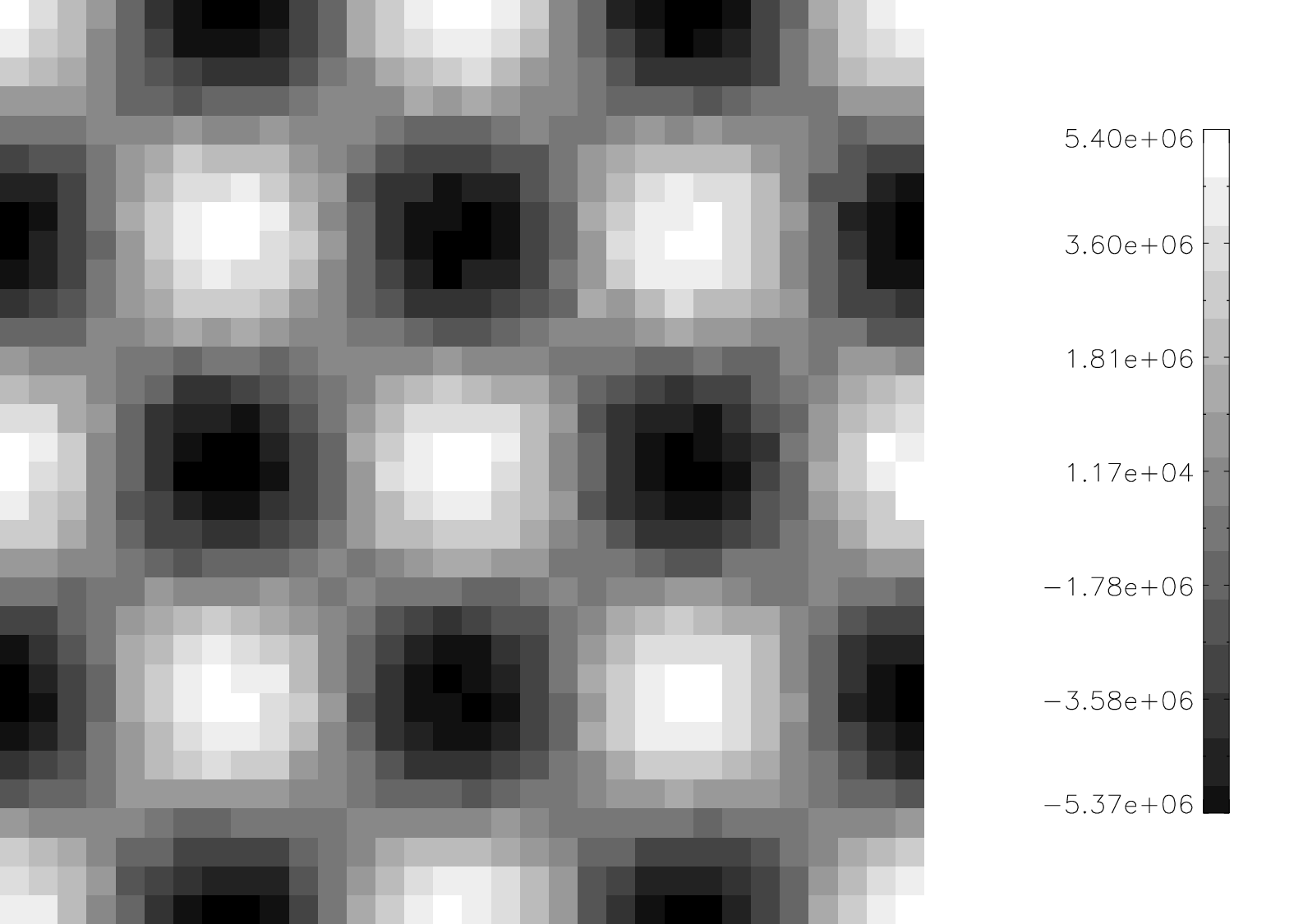}
\end{center}
\begin{center}
\includegraphics[width=5.0cm]{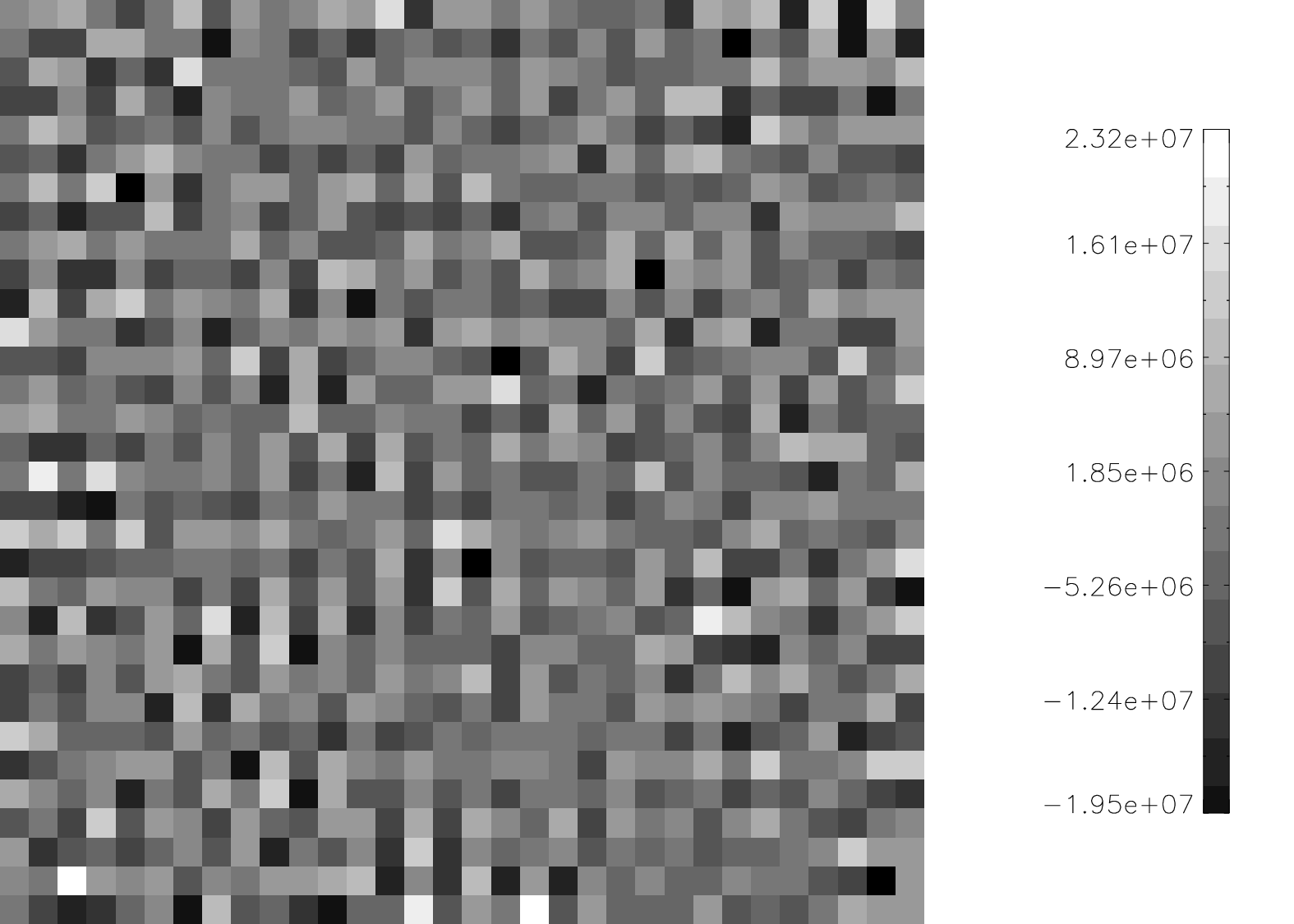}
\hspace{-0.0cm}\includegraphics[width=5.0cm]{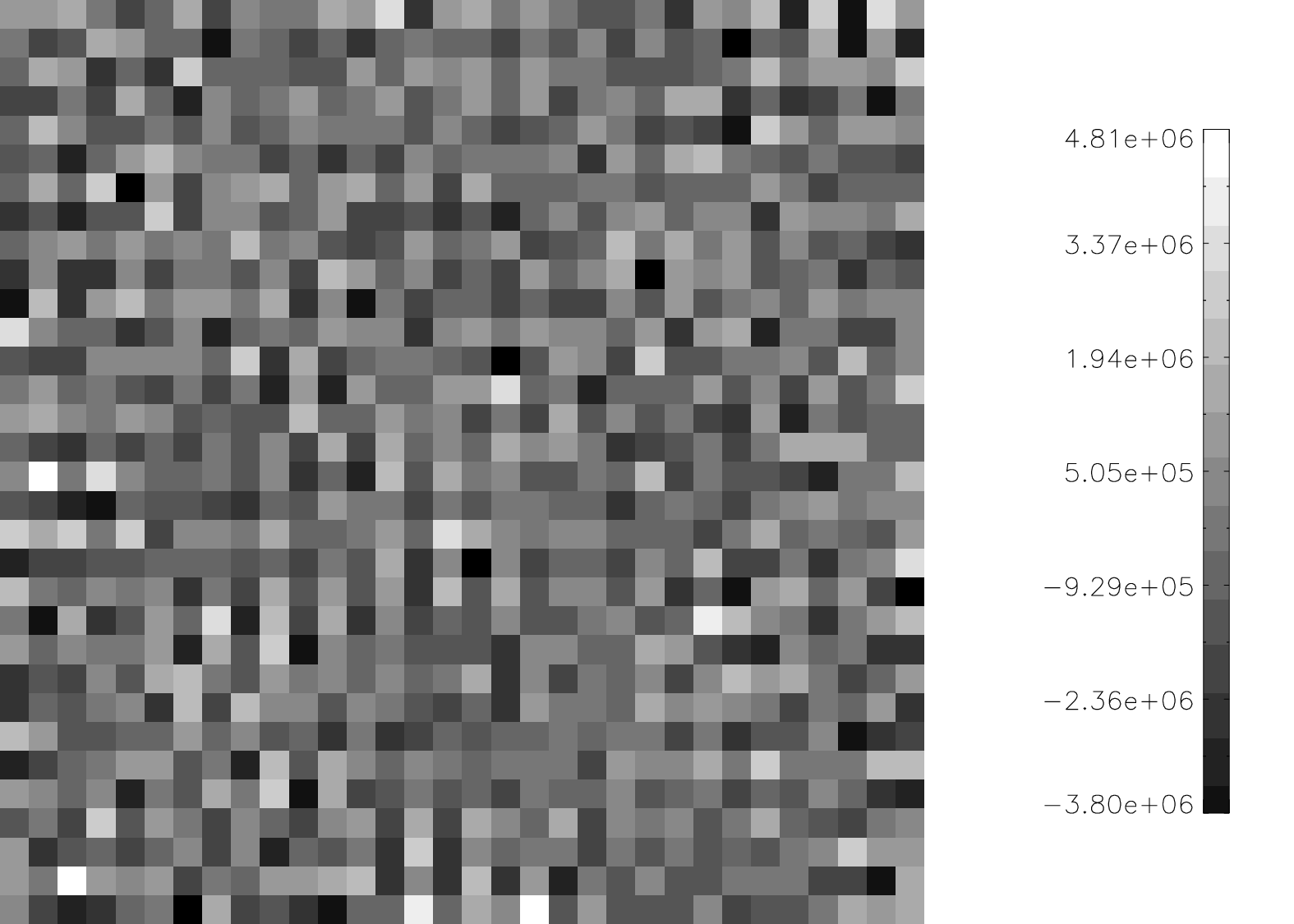}
\hspace{-0.0cm}\includegraphics[width=5.0cm]{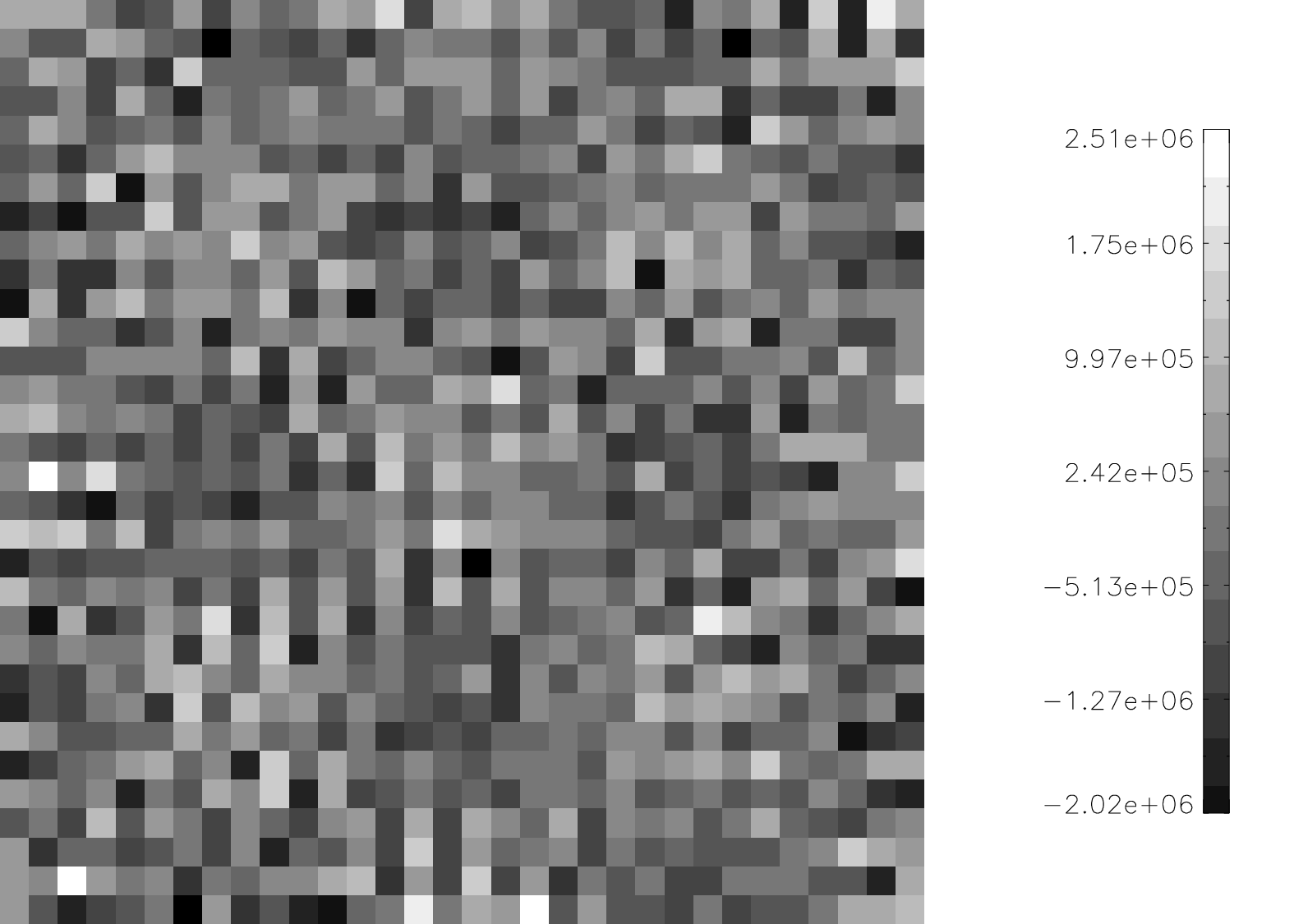} \caption{The
same as Figure 3, but for sample $B$.}
\end{center}
\end{figure}

As mentioned in \S 3.1, with $\psi_{E}$ and $\psi_{B}$ one can
produce the maps of $Q$ and $U$ by eqs.(33) and (34). Using
noise-added maps of $Q$ and $U$, we can further calculate noisy
variables $\mathbb{E}_{l_1,l_2}$ and $\mathbb{B}_{l_1,l_2}$ with
eqs.(16) and (17). This is the recovered maps of $\mathbb{E}_{l_1,l_2}$
and $\mathbb{B}_{l_1,l_2}$. Figures 2 and 3 present the recovered
maps of $\mathbb{E}_{l_1,l_2}$ and $\mathbb{B}_{l_1,l_2}$ for
samples $A$ and $B$, respectively. The signal-to-noise ratios (S/N)
are S/N=10, 50 and 100 from left to right. Comparing Figures 2 and 3
with Figure 1, we can conclude that the spatial structures of $E$
mode maps can be well recovered with the noisy maps $Q$ and $U$ if
$S/N\geq 10$. The recovery of $B$ mode structures is relatively
poor. For sample $A$ (Fig. 2), we may pick up the original
structures of $B$ field with all S/N$\geq$ 10 noisy maps of $Q$ and
$U$, while for sample $B$, the structures of $B$ field can not be
seen with $S/N=10$ map. That is, the structure identification of
sample $A$ is much better than sample $B$. This is because the field
of sample $A$ is highly inhomogeneous, while sample $B$ is not so
inhomogeneous. The latter is easily contaminated with a
statistically homogeneous Gaussian field.

\subsection{Recovery of power spectrum}
\begin{figure}[htb]
\begin{center}
\includegraphics[width=5.0cm]{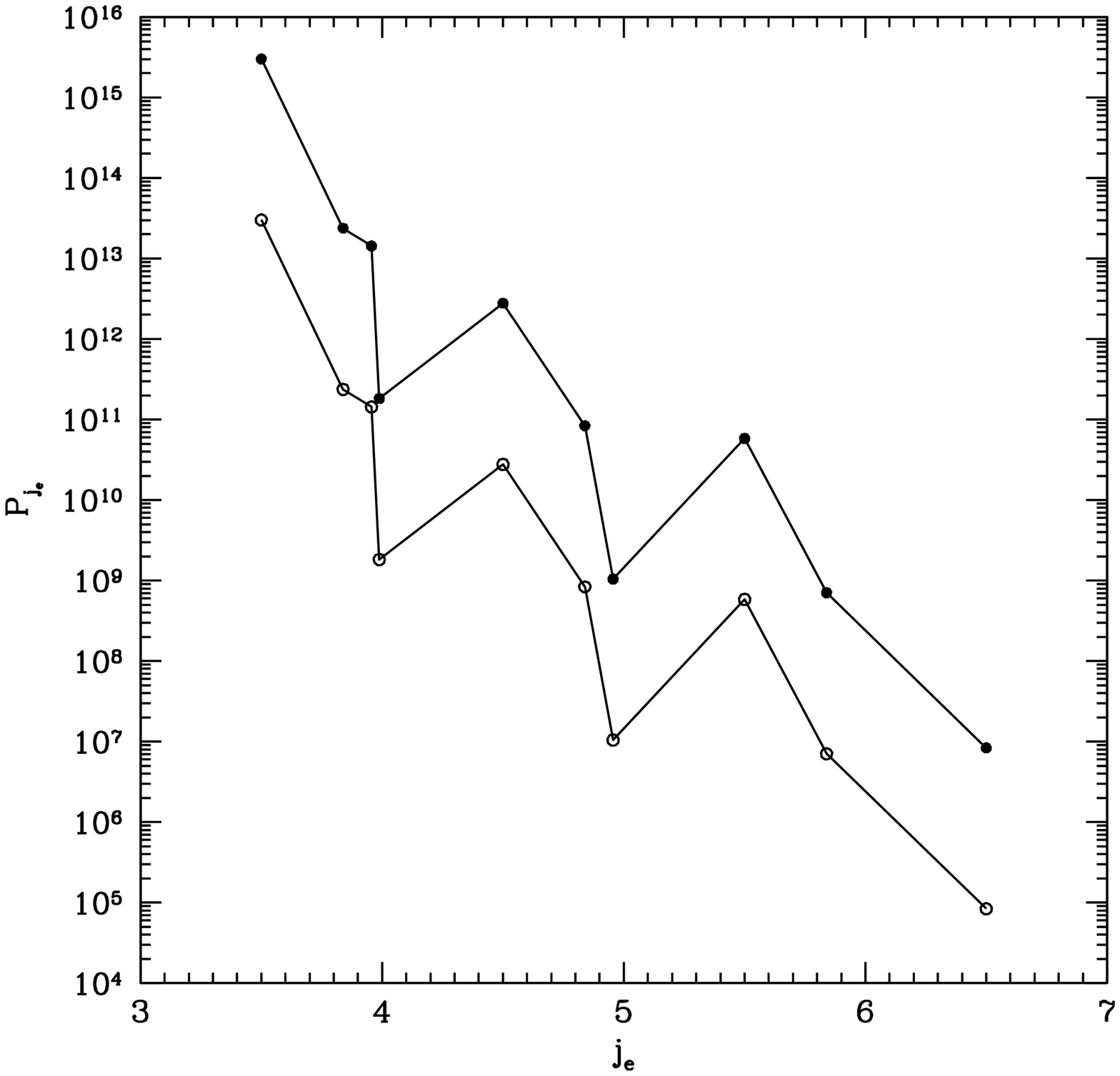}
\includegraphics[width=5.0cm]{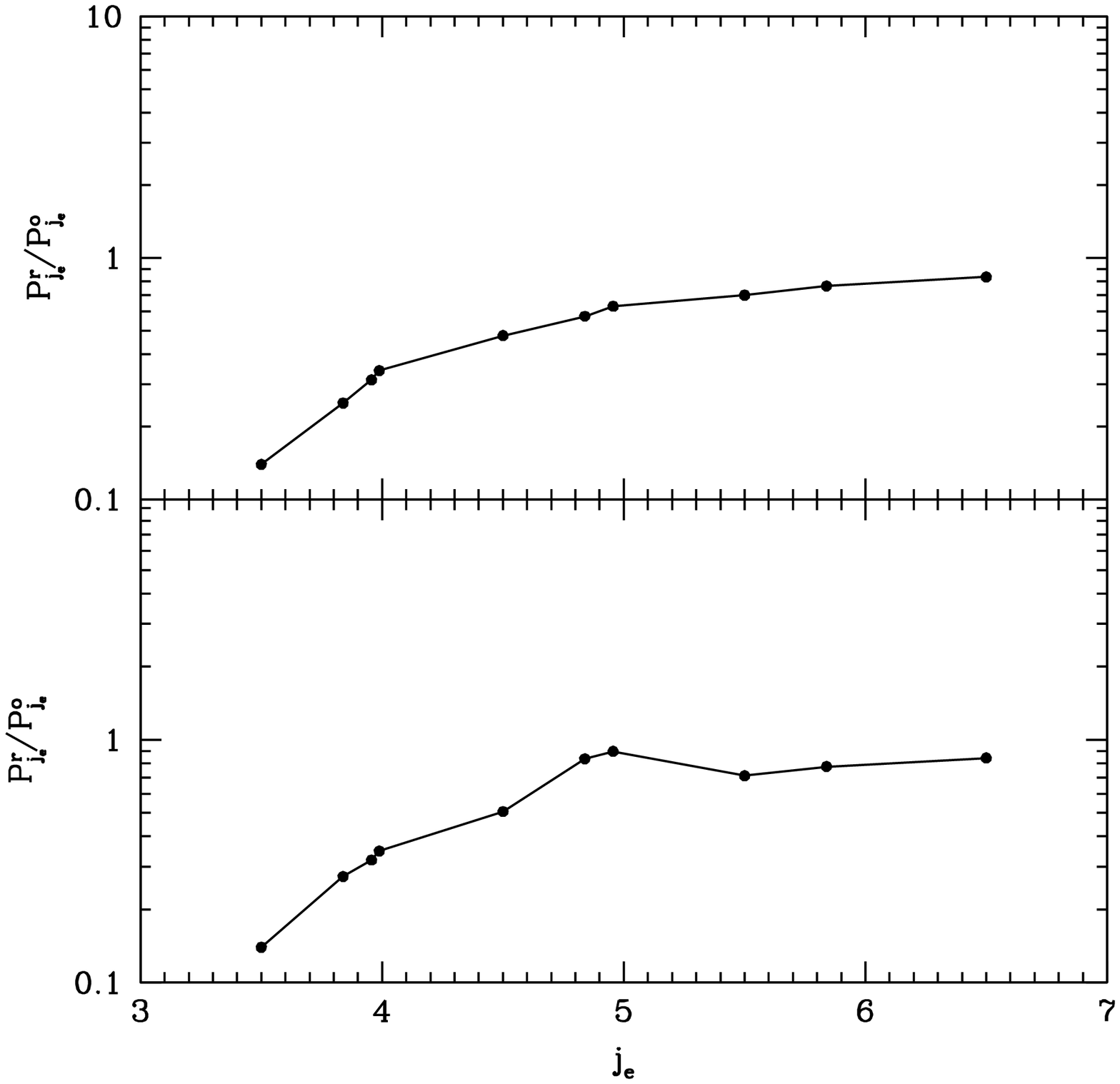}
\includegraphics[width=5.0cm]{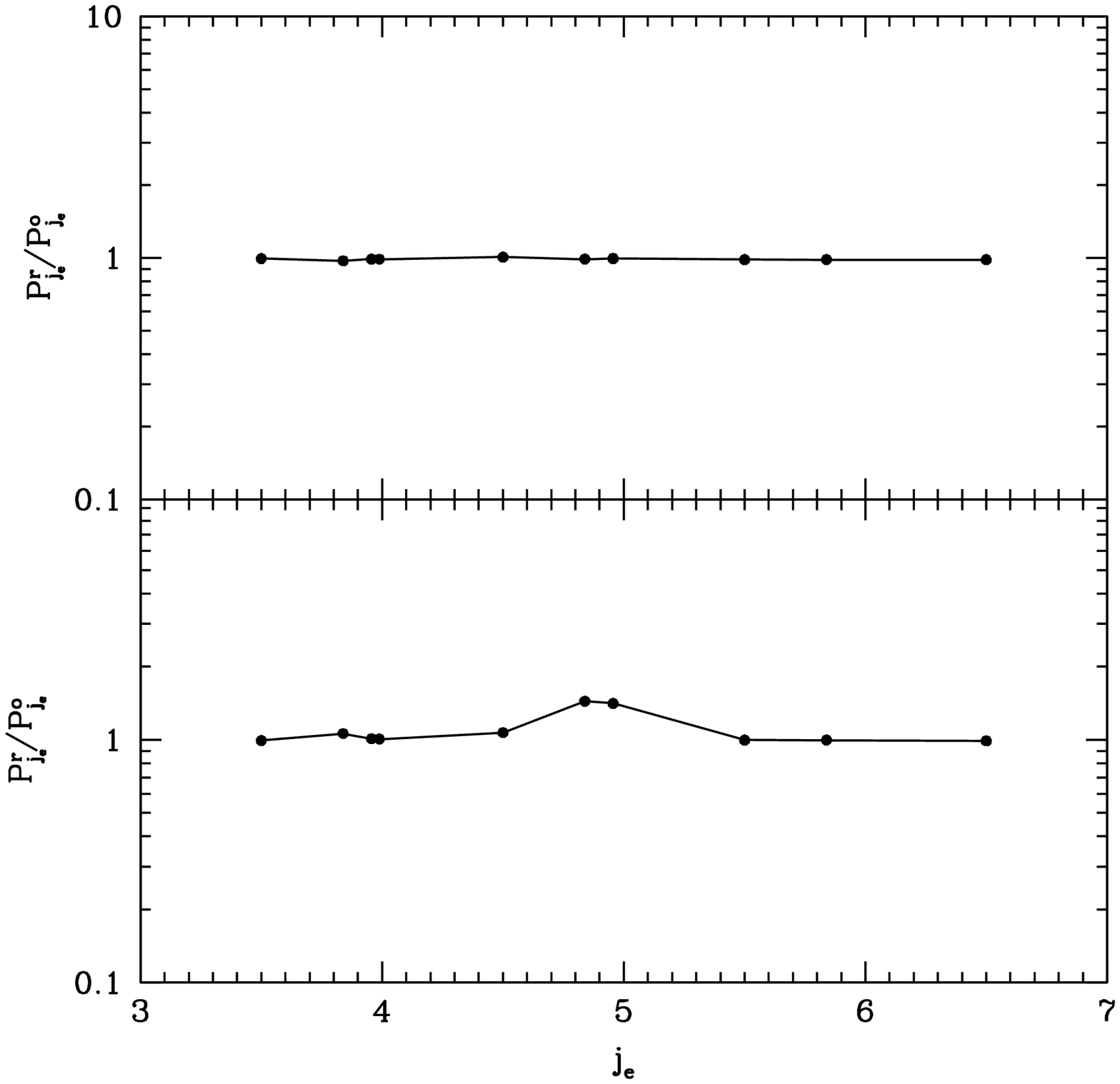}
\caption{Left panel: The DWT power spectra $P_{\bf j}$ of the
original maps $\mathbb{E}_{\bf l}$ and $\mathbb{B}_{\bf l}$ of
sample $A$. The power spectrum on the top is for the $E$ mode, and the
lower is for the $B$ mode. The ten data points correspond to $(j_1,
j_2)= $ (4,4), (4,5), (4,6), (4,7), (5,5), (5,6), (5,7), (6,6),
(6,7) and (7,7) from left to right. The power ratio $E/B$ is equal to
$10^2$. Middle panel: the ratio $P^{o}_{\bf j}/P^{r}_{\bf j}$,
where $P^{o}_{\bf j}$ is the original power spectra from maps
$\mathbb{E}_{l_1,l_2}$ and $\mathbb{B}_{l_1,l_2}$ of eq.(32), and
$P^{r}_{\bf j}$ is the recovered power spectra of $Q$ and $U$ with
eqs.(16) and (17) and without dropping boundary cells. The top is
for $E$ mode, and  bottom is for $B$ mode. Right panel: the same
as middle panel, but with 4 boundary cells dropped.}
\end{center}
\end{figure}
\begin{figure}[htb]
\begin{center}
\includegraphics[width=5.0cm]{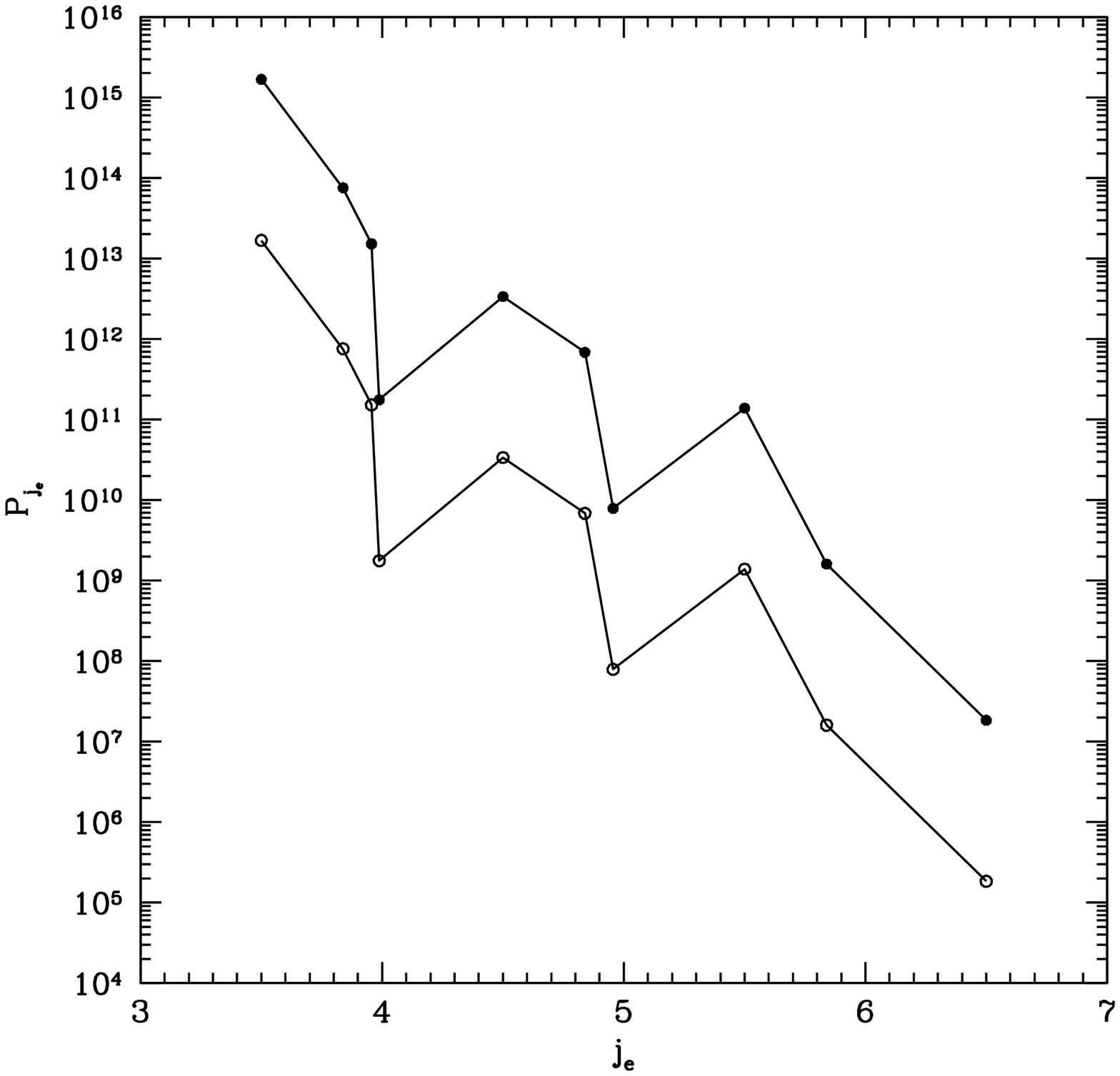}
\includegraphics[width=5.0cm]{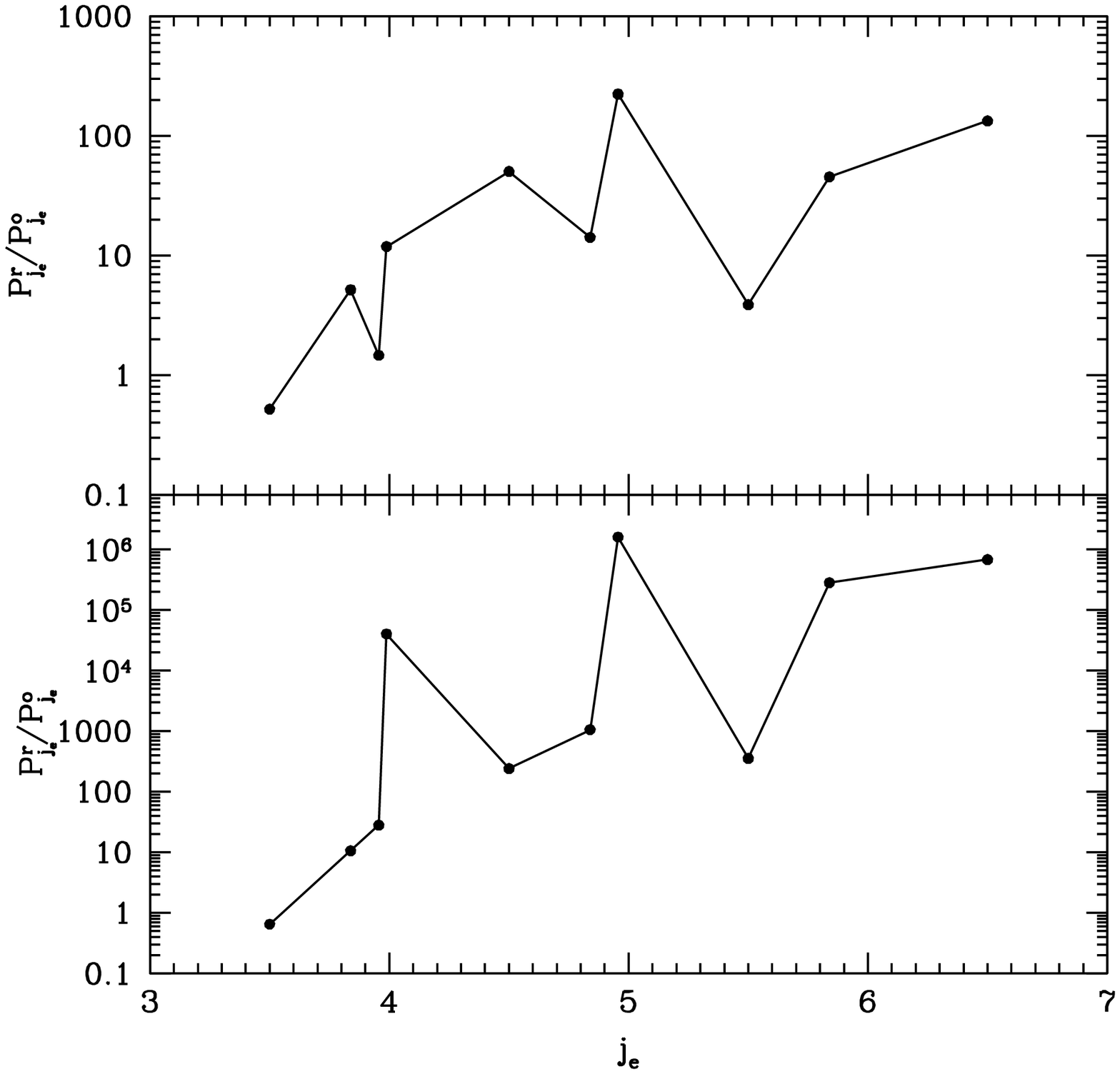}
\includegraphics[width=5.0cm]{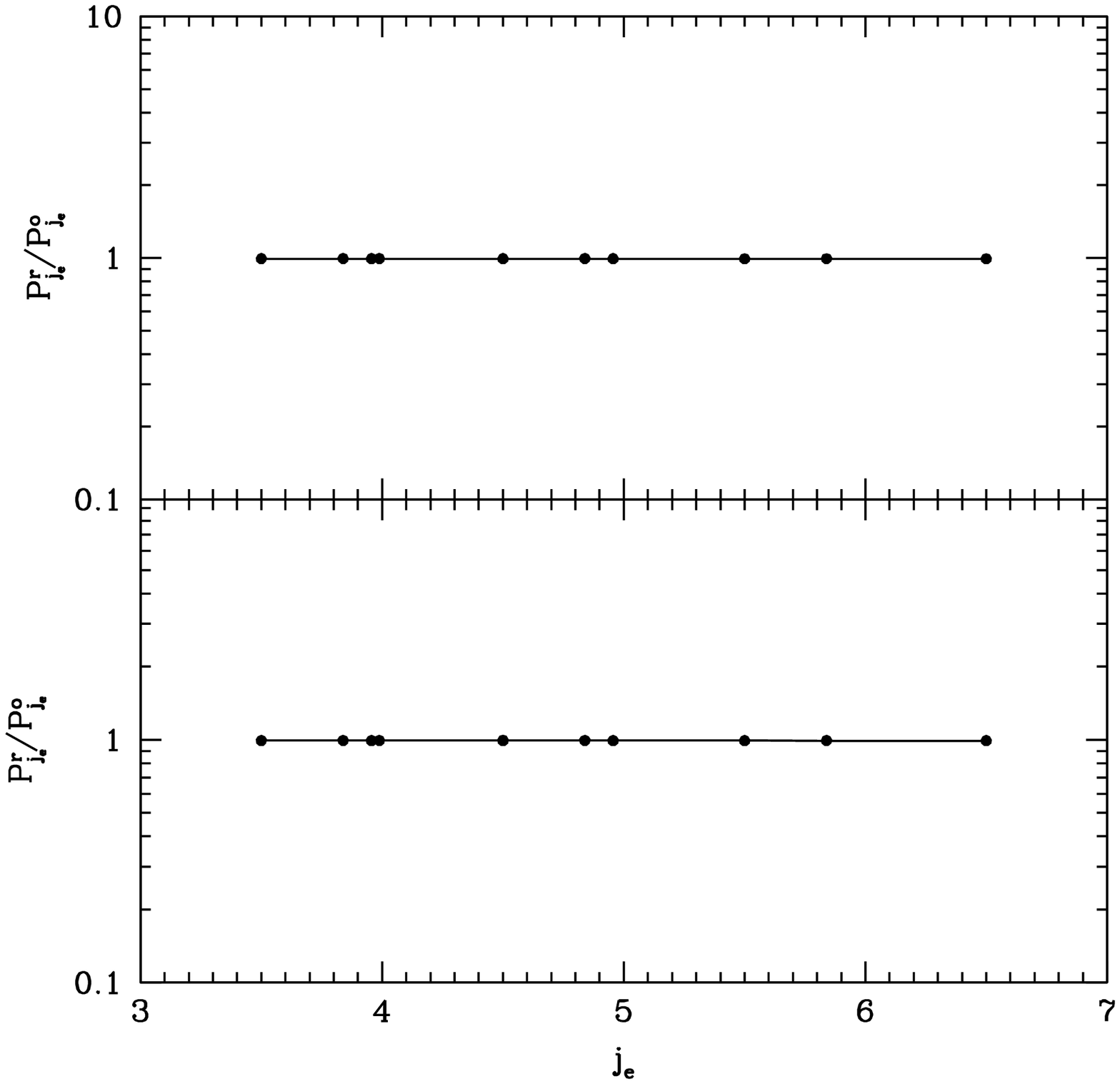}
\caption{The same as Figure 4, but for sample $B$.}
\end{center}
\end{figure}

We now turn to the recovery of the power spectrum. For wavelet
Daubechies 6, the non-zero elements of the
matrix $T^n_{l,l'}$ are in a band $|l-l'|\leq 4$, and therefore,
cells distant from boundary, larger than $\Delta l =4$, will be less
affected by the boundary. Thus, one may expect that the power
spectrum recovery would be reasonable with dropping 4 boundary
cells.

Figures 4 and 5 show the DWT power spectra of $\mathbb{E}_{l,l'}$ and
$\mathbb{B}_{l,l'}$ of both original and recovered samples of sets
$A$ and $B$, respectively. It includes 1.) the power spectra of the
original maps, i.e. the map directly given by $E$ and $B$ [eq.(32)];
2.) the ratio $P^{o}_{\bf j}/P^{r}_{\bf j}$, where $P^{o}_{\bf j}$
is the original power spectra from maps $\mathbb{E}_{l_1,l_2}$ and
$\mathbb{B}_{l_1,l_2}$ of eq.(32), and $P^{r}_{\bf j}$ is the
recovered power spectra of maps $Q$ and $U$ with eqs.(16) and (17)
without dropping boundary cells; 3.) the same as 2.) but dropping 4
boundary cells in the recovered power spectra.

In Figures 4 and 5, the scale ${\bf j}=(j_1,j_2)$ is described by an effective
scale defined as $j_{\rm eff}$ as $(1/2^{j_{\rm eff}})= [ (1/2^{j_1})^2+
(1/2^{j_2})^2]^{1/2}$. The samples are symmetric with respect to $x
\rightleftharpoons y$. The power of mode $(j_1,j_2)$ should be the
same as $(j_2,j_1)$. Thus, for $J=8$, the available pairs
$(j_2,j_1)$ are (4,4), (4,5), (4,6), (4,7), (5,5), (5,6), (5,7),
(6,6), (6,7) and (7,7), corresponding to $j_{\rm eff}=3.50$, 3.84, 3.95,
3.98, 4.50, 4.84, 4.96, 5.50, 5.84, 6.50. The modes on scales with $j_1,
j_2\leq 3$ are dropped, as all cells are affected by the boundary
effect.

 We see from Figures 4 and 5 that the power spectra can indeed
be well recovered by dropping 4 boundary cells. However, the
boundary effect of sample $A$ are less serious than sample $B$. This
is because,  for sample $A$, both $\psi_{E,B}(x,y)$ and
$\partial_n\psi_{E,B}(x,y)$ are very small at boundary. The
contribution to power by boundary cells is low. On the other hand,
for sample $B$, the power spectra recovered without dropping
boundary cells are significantly different from the original one.
The $B$-mode power spectrum is hugely affected by the boundary. On
small scale the derivative operator in the DWT representation is
determined by data at a few discrete points, which leads to large
error. This problem is always present in algorithms involving taking
derivative on discrete data sets. Nevertheless the error caused by
boundary is decreases rapidly as the scale increases.
\begin{figure}[htb]
\begin{center}
\includegraphics[width=5.0cm]{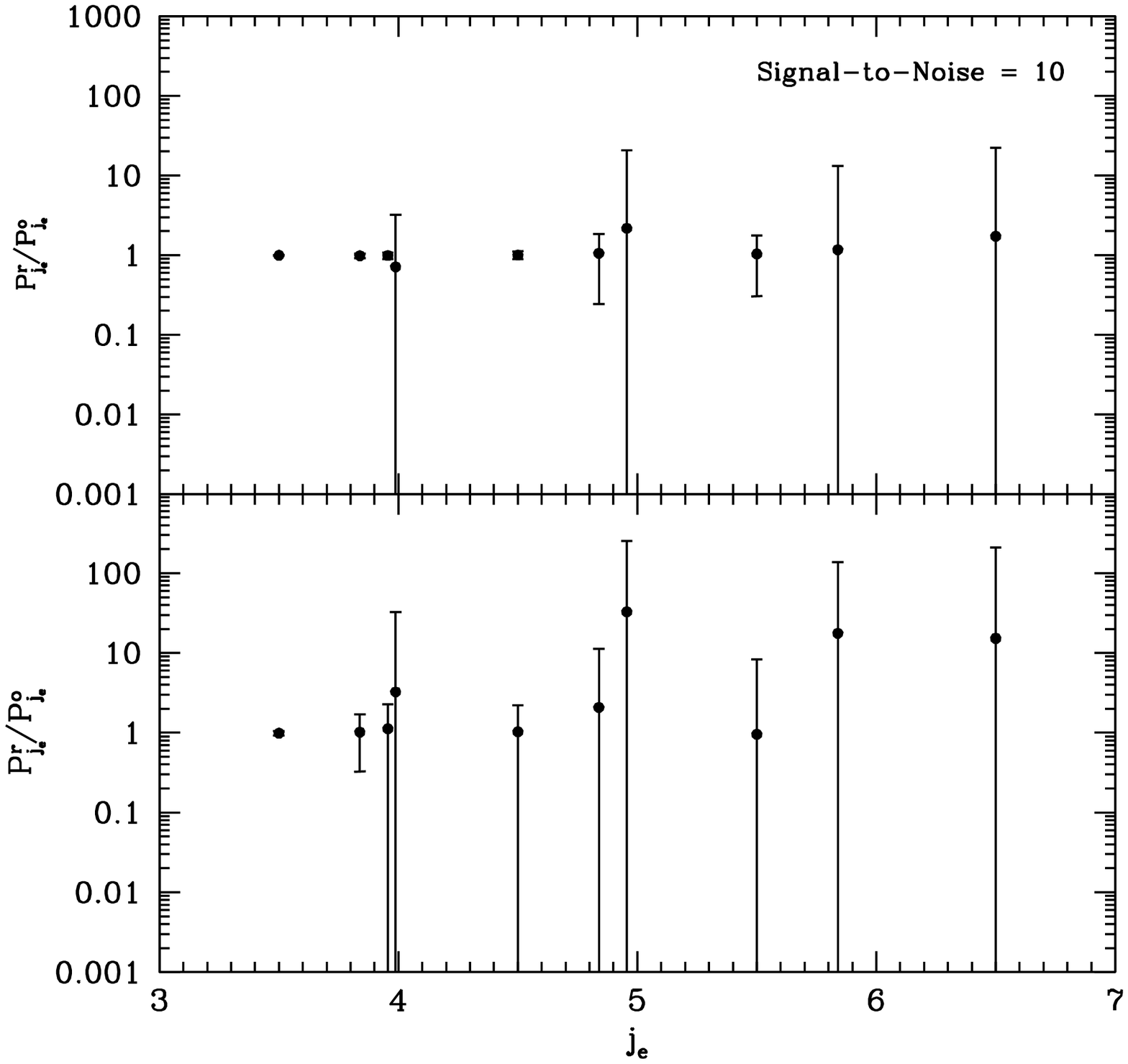}
\includegraphics[width=5.0cm]{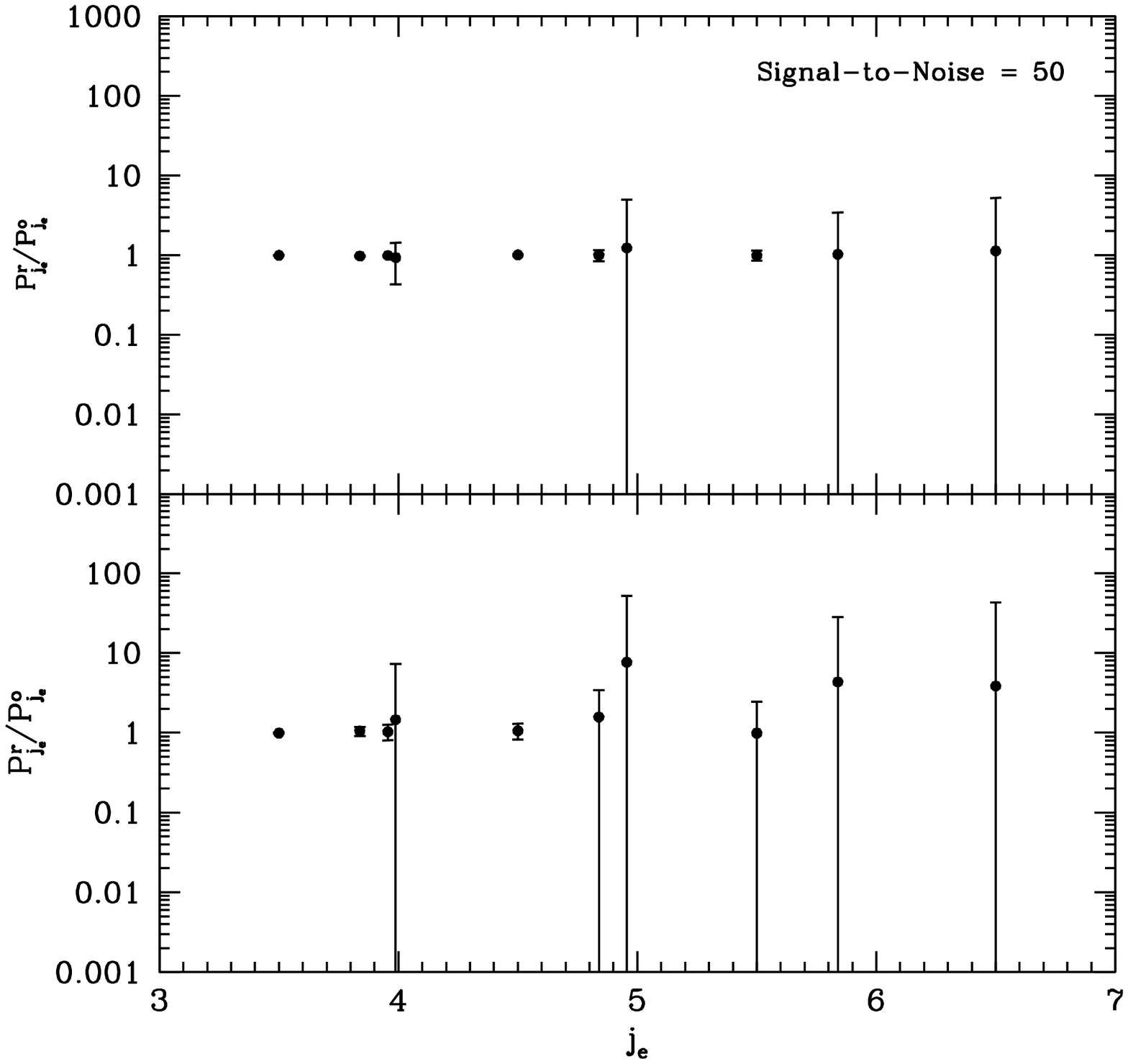}
\includegraphics[width=5.0cm]{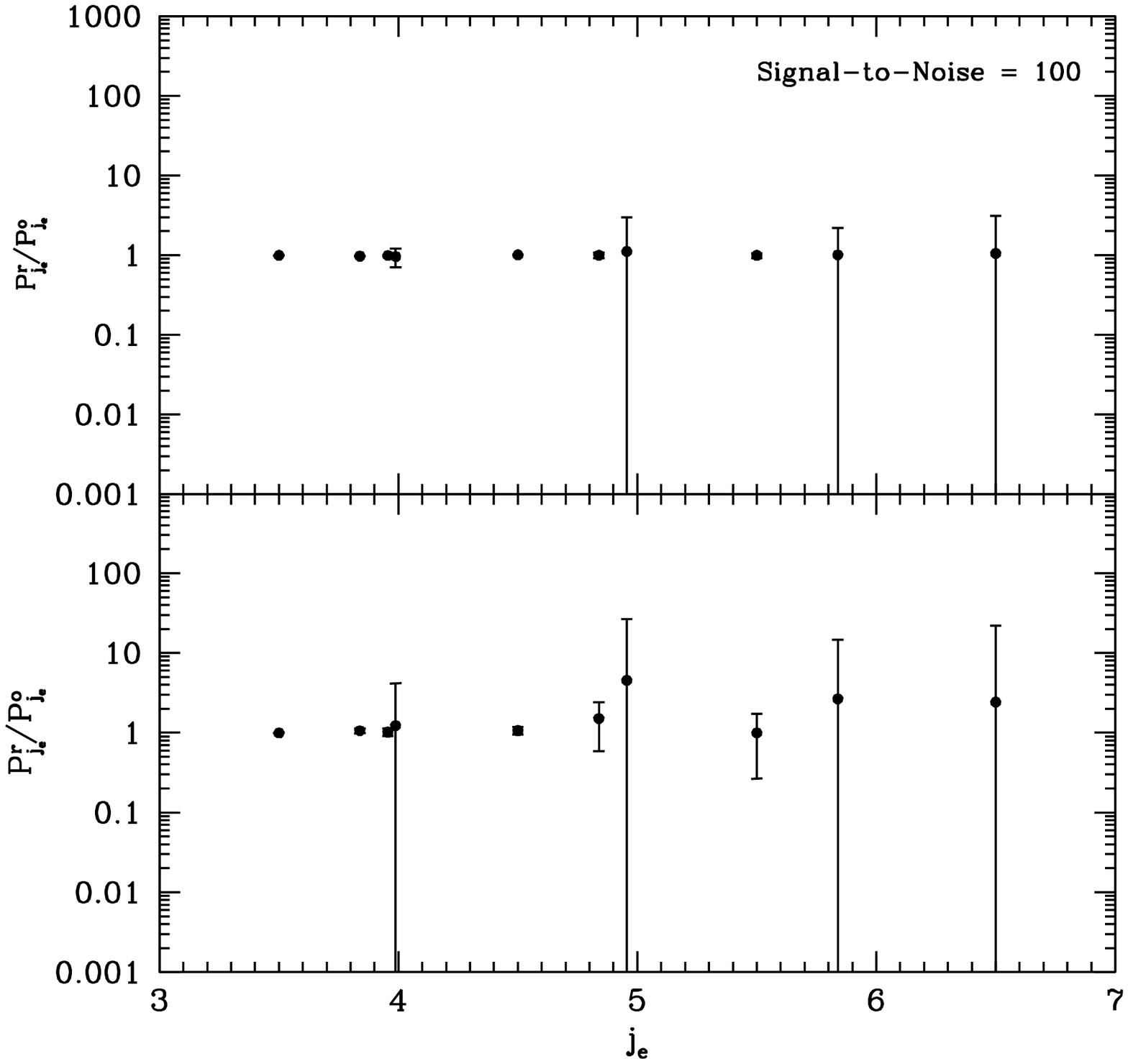}
\caption{The ratio $P^{o}_j/P_j^{r}$ for sample $A$, where
$P^{o}_j$ is the power spectrum of the original maps
$\mathbb{E}_{l_1,l_2}$ and $\mathbb{B}_{l_1,l_2}$, and $P_j^{r}$
is calculated by eqs.(30) and (31) from $Q$ and $U$ with noise
addition on the level of S/N equal to 10 (left), 50(middle), and
100(right). In each panel,the top is for $E$ mode, and the bottom
is for $B$ mode.}
\end{center}
\end{figure}
\begin{figure}[htb]
\begin{center}
\includegraphics[width=5.0cm]{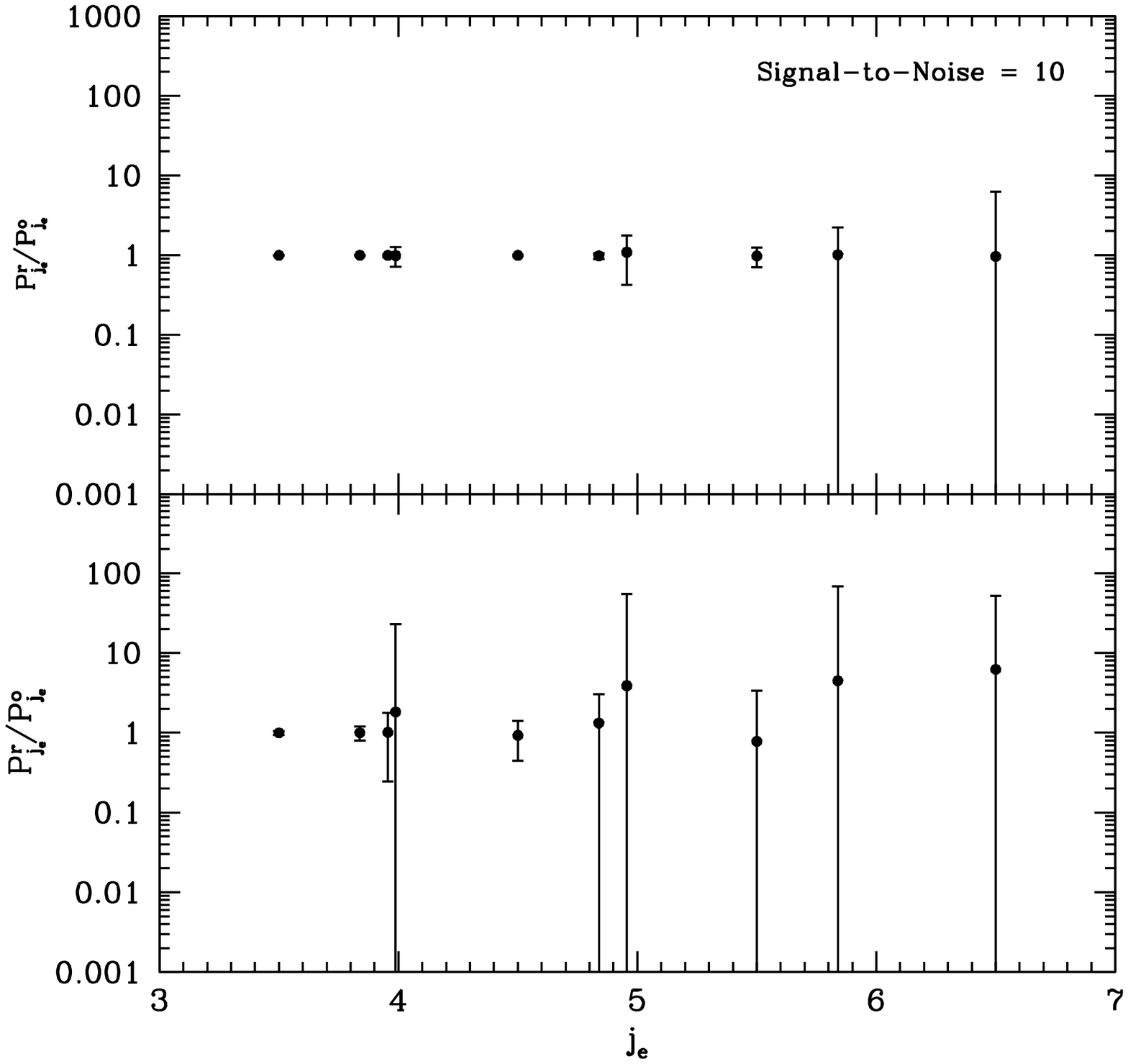}
\includegraphics[width=5.0cm]{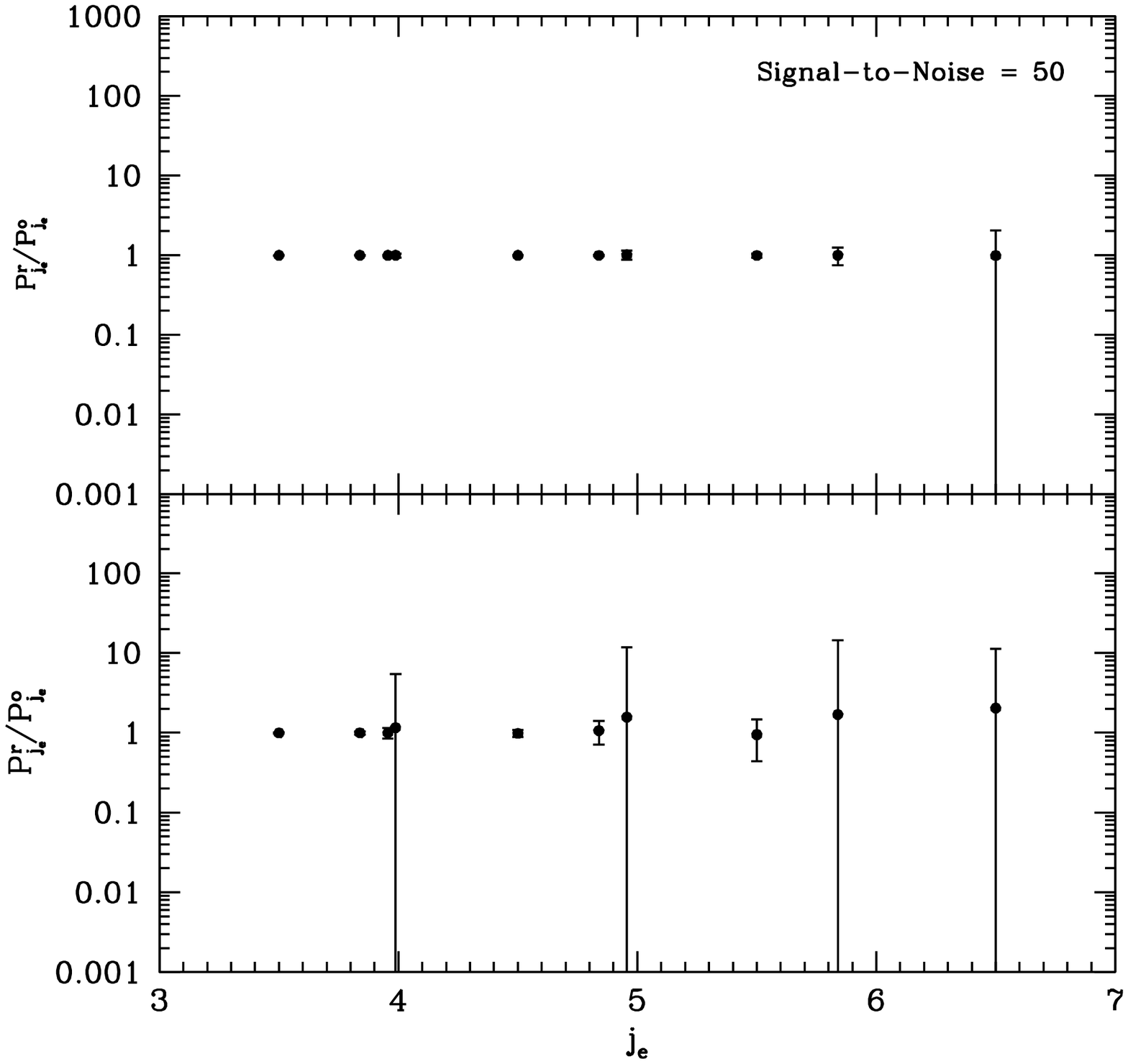}
\includegraphics[width=5.0cm]{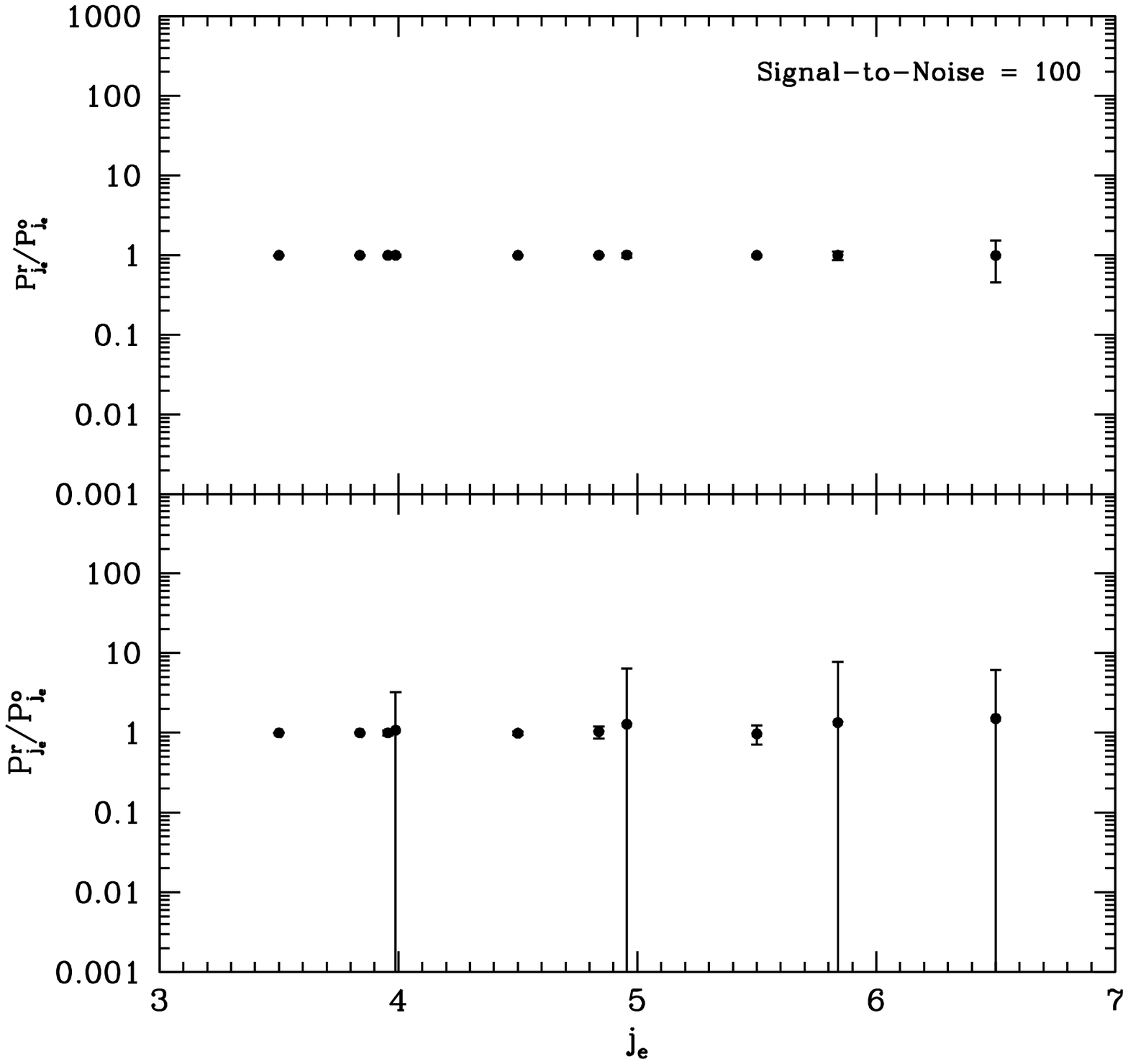}
\caption{The same as figure 6, but for sample $B$.}
\end{center}
\end{figure}

To measure how good the recovery of power spectrum is, we use the
ratio $P^{o}_{\bf j}/P^{r}_{\bf j}$ to describe the deviation of the
recovered power spectrum with noisy maps from the original one. All
error bars are the variances calculated from 100 independent noisy
maps. The results are plotted in Figures 6 and 7.

First we see that the effect of Gaussian noise is small at larger
scales, because the noise is added on each pixels (finest scale) of
the maps $Q$ and $U$, and the uncertainty on large scales is
suppressed. This point can also be seen from the fact that the error
bars of modes (4,7), (5,7),(6,7) and (7,7) are much larger than
others. It is because the Gaussian noise on smallest scales, $j_1$
or $j_2=7$, is not suppressed.

Figures 6 and 7 show that other than the modes with $j_1$ or
$j_2=7$, the power of $E$ mode can be reasonably recovered up to
mode (6,6), or $J_{\rm eff}=5.5$, when S/N$=$ 10. As expected, the
recovery for $B$ mode generally is poor. Nevertheless, we can
recover the DWT powers of $B$ mode till (5,5), or $J_{\rm
eff}=4.5$, when S/N=50 (sample A) or S/N=10 (sample B).

An interesting point shown in Figures 6 and 7 is that the effects
of noise on samples $A$ and $B$ are different. The error bars of
sample $A$ generally are larger than that of sample $B$. This is
probably because for sample $A$, other than the central part, most
cells are smooth, and have low local fluctuations. For those
cells, the fluctuations of noise will strongly contaminate
the power of original field, especially when derivative is
involved. On the other hand, for sample $B$, most cells have
relatively stronger local fluctuations, and the effect of noise is
relatively low.

\section{Tests with samples of Gaussian random fields}

\subsection{Samples}

With the preparation given in the previous section, we can consider
the case that $\psi_E(x,y)$ and $\psi_B(x,y)$ as random fields. The
sample of $E$ and $B$ can still be generated with the same procedure
of \S 3.1, but $\psi_E(x,y)$ and $\psi_B(x,y)$ are taken to be
Gaussian random fields with Fourier power spectra $a_E k^{-\alpha}$
and $a_Bk^{-\alpha}$, respectively, and $\alpha=3.6$. The variable
$k^2=k_x^2+k_y^2$, and $k_x$, $k_y$ are the Fourier variables of
$x$, $y$ space, respectively. The constant factors $a_E$ and $a_B$
are used to adjust the ratio of $E/B$ power. From eq.(32), the power
spectra of $E$ and $B$ are
\begin{equation}
P_E=a_Ek^{2-\alpha}, \hspace{5mm} P_B=a_B k^{2-\alpha}.
\end{equation}
We produce the maps in an area described by coordinate $(x,y)$ in
range $-0.5\leq x\leq 0.5$ and $-0.5\leq y\leq 0.5$ with pixelized
into 512$\times$512. With maps $E$ and $B$, one can find the maps
$\mathbb{E}$ and $\mathbb{B}$ by eqs.(13), and (14).

\begin{figure}[htb]
\begin{center}
\includegraphics[width=7.5cm]{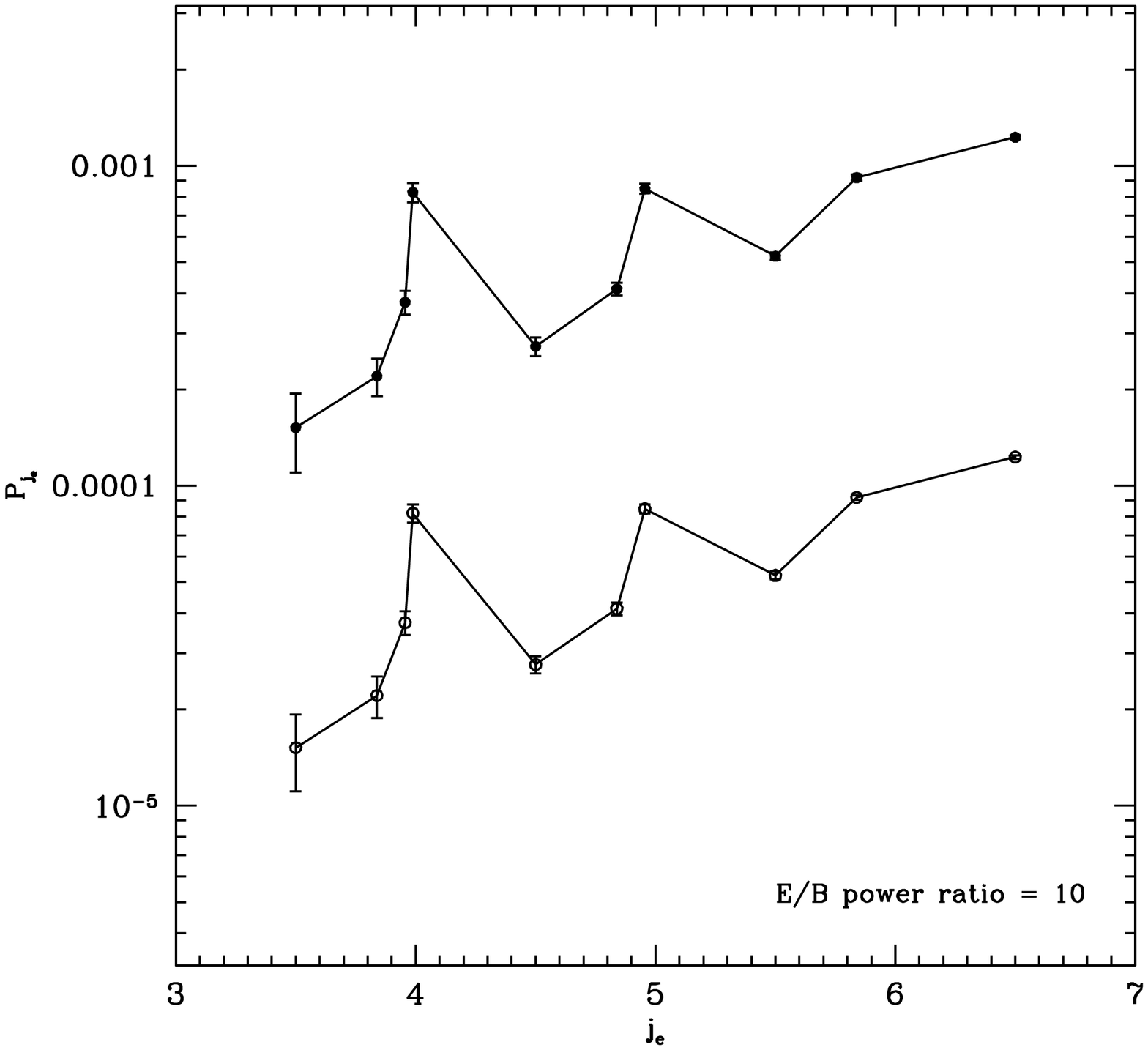}
\end{center}
\vspace{-1cm} \caption{The power spectra of $\mathbb{E}$ (higher
one), $\mathbb{B}$ (lower one), for which $\psi_E$ and $\psi_B$ are
Gaussian random field with the Fourier power spectra $a_E
k^{-3.6}$ and $a_Bk^{-3.6}$, respectively. The ten data points
correspond to $(j_1, j_2)= $ (4,4), (4,5), (4,6), (4,7), (5,5),
(5,6), (5,7), (6,6), (6,7) and (7,7) from left to right. The error
bars are from 100 samples of the Gaussian random field $\psi_E(x,y)$
and $\psi_B(x,y)$. The ratio of powers of $E$ and d $B$ modes is
equal to 10.}
\end{figure}

The DWT power spectrum of $\mathbb{E}$ and $\mathbb{B}$ is shown in
Figure 8. Since $J=8$, the available modes $(j_1, j_2)$ still are
(4,4), (4,5), (4,6), (4,7), (5,5), (5,6), (5,7), (6,6), (6,7) and
(7,7). The modes with $j_1,j_2\leq 3$ are dropped, as they have only
boundary cells. The ratio of the E/B power is taken to be 10. The
error bars are from the variance of 100 samples. We see from Figure
8 that the powers of modes (4,7), (5,7) and (6,7) are nearly about
the same as (7,7). Similarly, the powers (4,6), (5,6) are nearly
about the same as (6,6); the power (4,5) is nearly about the same as
(5,5). It is because in the case of $j_1<j_2$, the power is
dominated by the small scale $j_2$.

\subsection{Effects of random field}

Unlike the maps in \S 3, all the maps of $\psi_E(x,y)$,
$\psi_B(x,y)$; $Q$, $U$; and $\mathbb{E}$, $\mathbb{B}$ are random
fields. A serious problem caused by random fields is that the power
of $E$ mode may leak to $B$ mode, and vice versa. That is, even when the
original $B$ mode power is zero, the recovered $B$ mode power would not
be zero.

\begin{figure}[htb]
\begin{center}
\includegraphics[width=7.5cm]{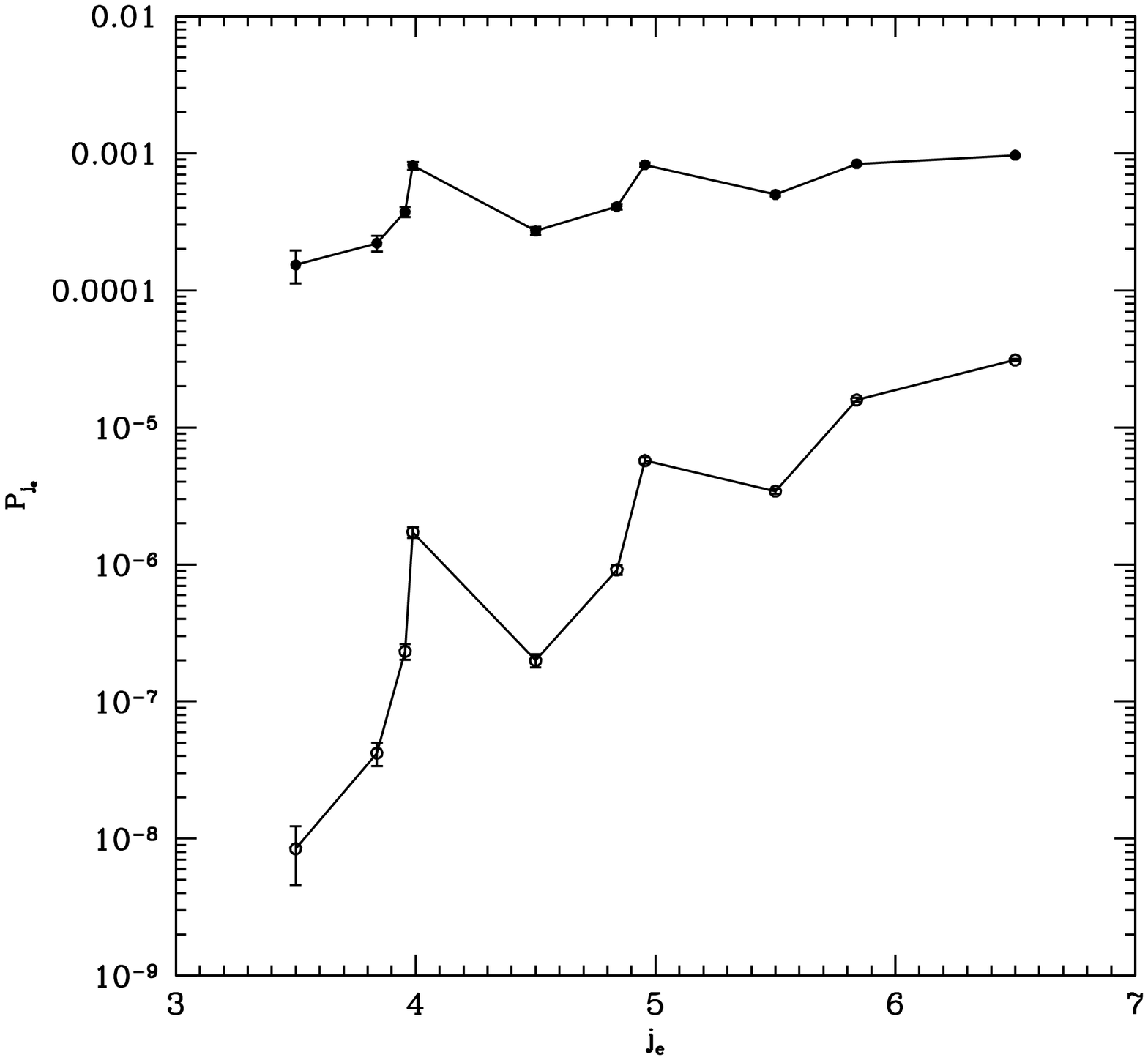}
\end{center}
\vspace{-1cm} \caption{DWT power spectra of $E$ mode (top) and $B$
mode (bottom). The $\psi_E(x,y)$ is a Gaussian fields with the
same Fourier power spectrum as Figure 8, while taking
$\psi_B(x,y)=0$.}
\end{figure}
\begin{figure}[htb]
\begin{center}
\includegraphics[width=7.5cm]{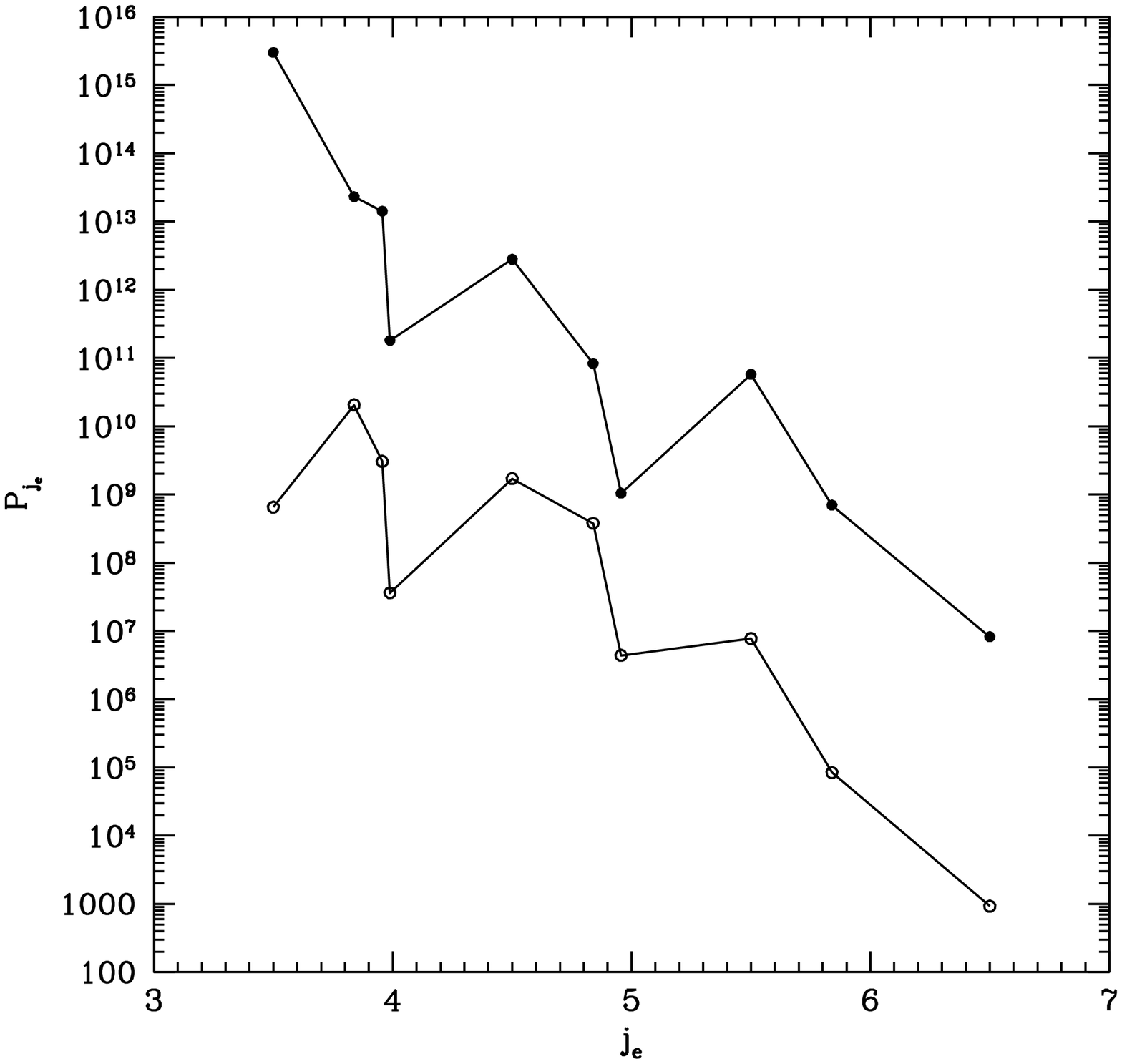}
\includegraphics[width=7.5cm]{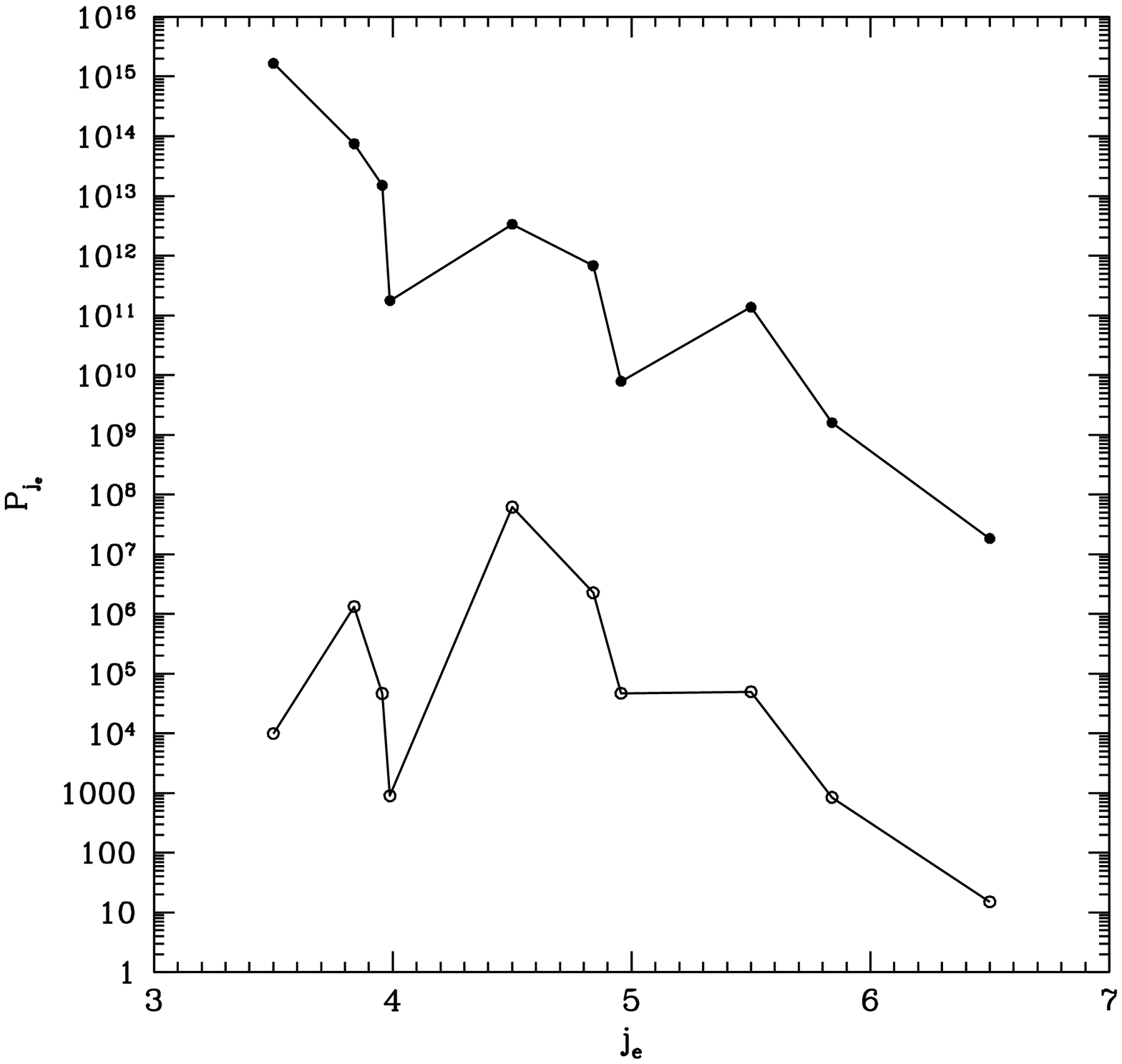}
\end{center}
\vspace{-1cm} \caption{DWT power spectra of $E$ mode (top) and $B$
mode (bottom). The $\psi_E(x,y)$ is given by sample A and sample B,
while taking $\psi_B(x,y)=0$. The left panel is for sample A, while
the right panel is for sample B.}
\end{figure}

To demostrate the power leakage, we take $\psi_E(x,y)$ to be a Gaussian
random field with the same Fourier power spectrum as Figure 8, while
$\psi_B(x,y)$ to be 0. That is, the power of $B$ mode originally is
zero. Figure 9 presents the recovered DWT power spectra of $E$- and
$B$-modes. We see that the recovered $E$-mode power spectrum is
nearly the same as the original one shown in Figure 8, except that
the recovered power spectrum on the finest scale is a little smaller
than the original one. However, the recovered $B$-mode power
spectrum is not zero. It is spurious $B$ power. It arises from the
leaking of $E$ mode power to $B$ mode. On large scales $j_{\rm
eff}\leq 4$, the ratio of the $E/B$ power is about $10^4$, while on
small scales $j_{\rm eff}\geq 6$, this ratio is less than 10$^2$.
This is caused by the variance of random field. Thus, one may
conclude that for the $512\times 512$ sample of a Gaussian random
fields of eq.(35), the developed algorithm would be effective only
if the ratio $E/B$ is less than $10^2$ on small scales.

As a comparison, we plot Figure 10, in which the $\psi_E(x,y)$ is
given by samples A and B, while $\psi_B(x,y)=0$, i.e. the power of
$B$ mode originally is also zero. Figure 10 shows that the recovered
$B$-mode powers are also not zero. However, it generally is less
than the original one by at least 3 orders. That powers seem to come
from the numerical processes. Therefore, the errors caused by the variance
of random fields are serious.

\subsection{Recovery of $E$, $B$ power spectra}

As in \S 3.3, we measure the soundness of the recovery of
power spectrum by the ratio $P^{r}_{\bf j}/P^{o}_{\bf j}$, where
$P^{o}_{\bf j}$ is the power spectra of original maps
$\mathbb{E}_{l_1,l_2}$ and $\mathbb{B}_{l_1,l_2}$ from eq.(32),
and $P^{r}_{\bf j}$ is the recovered power spectra from noisy maps
of $Q$ and $U$. The Gaussian noise added on the maps $Q$ and $U$
are on the levels S/N=100, 20 and 10. Similar to \S 3.3, four
boundary cells are dropped. The results are plotted in Figures 11,
12 and 13, for which the ratio of the powers of $E$ and $B$ are
equal to, 10, 20 and 100, respectively.
\begin{figure}[htb]
\begin{center}
\includegraphics[width=5.0cm]{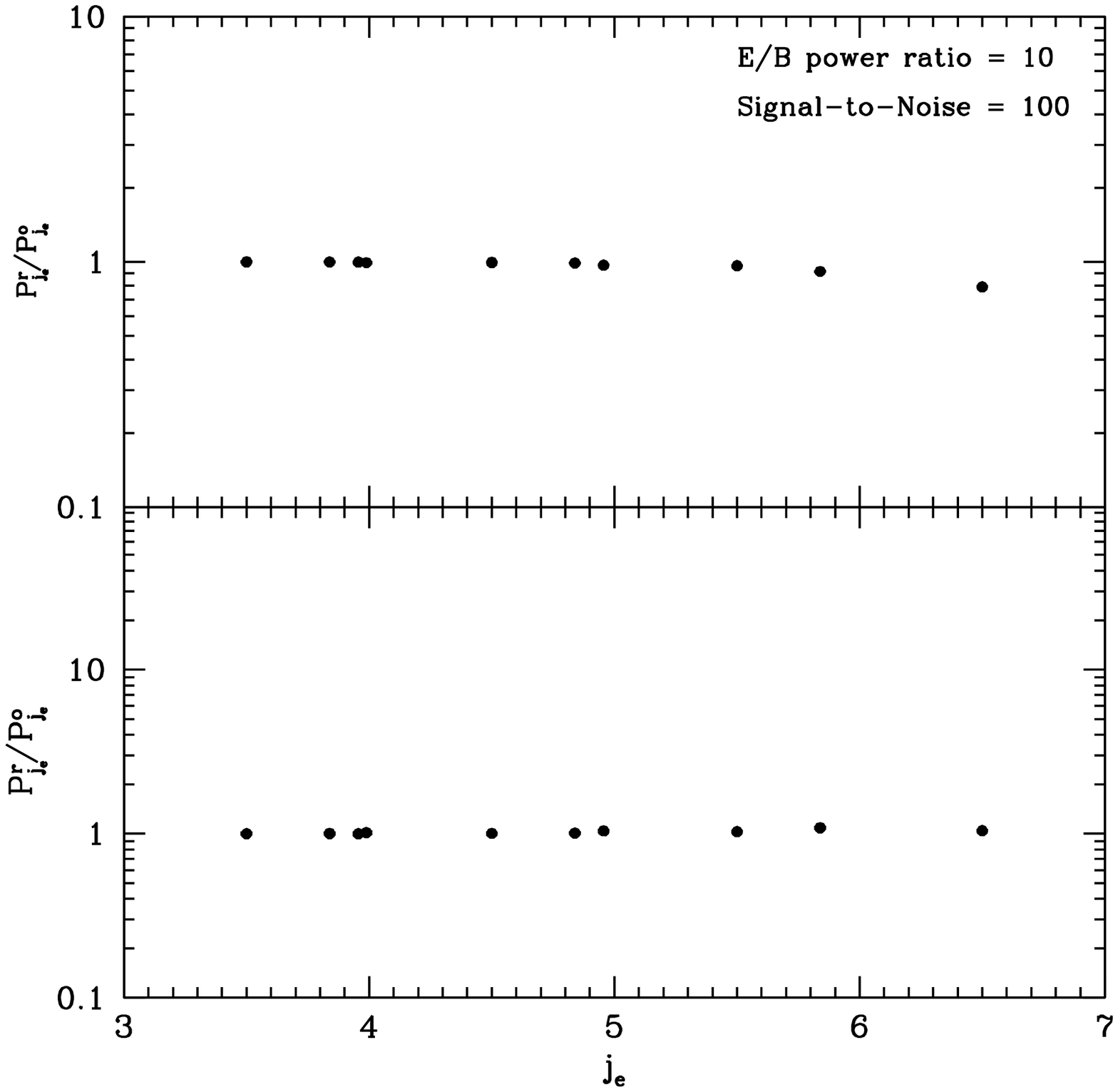}
\hspace{-0.0cm}\includegraphics[width=5.0cm]{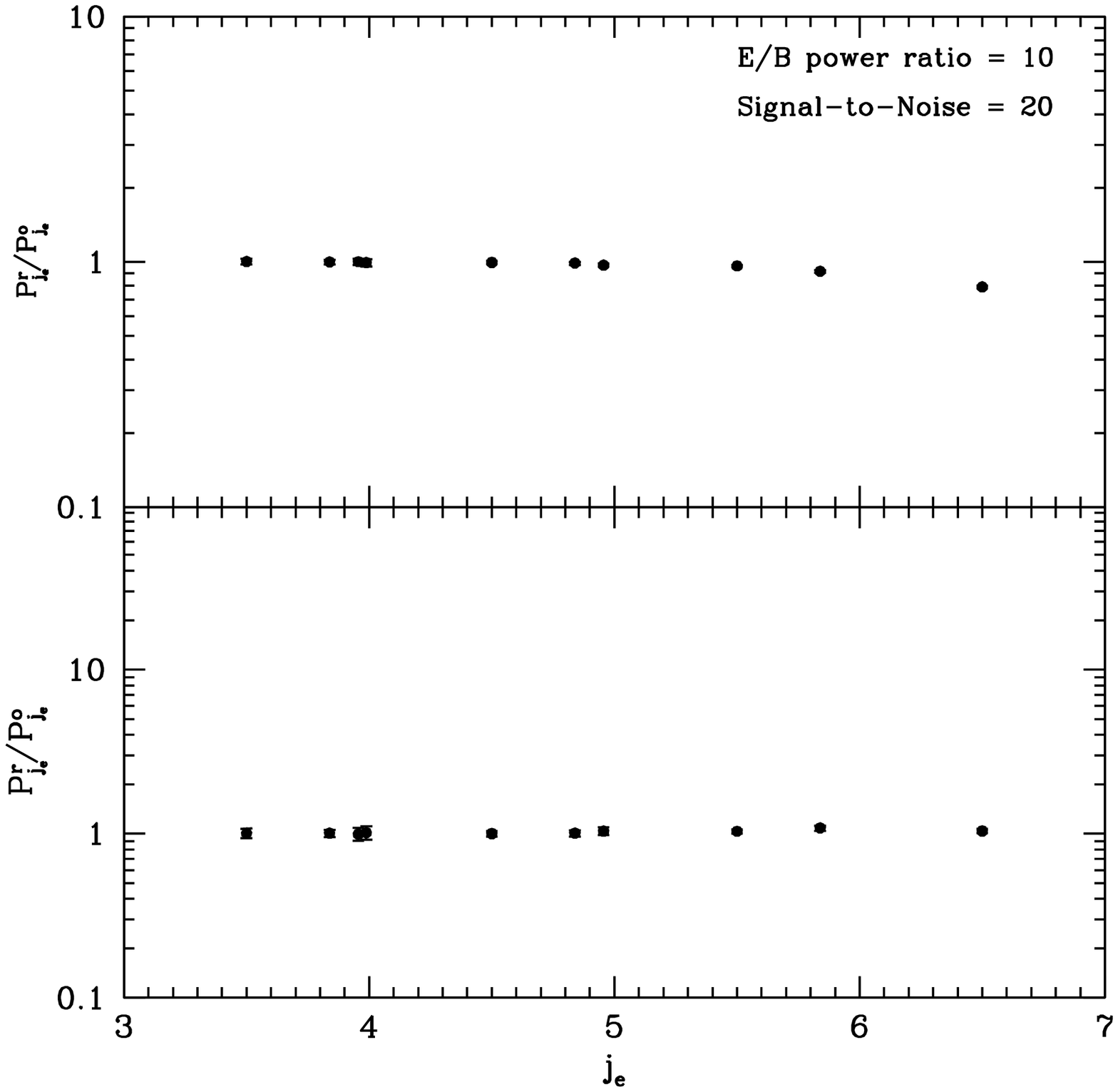}
\hspace{-0.0cm}\includegraphics[width=5.0cm]{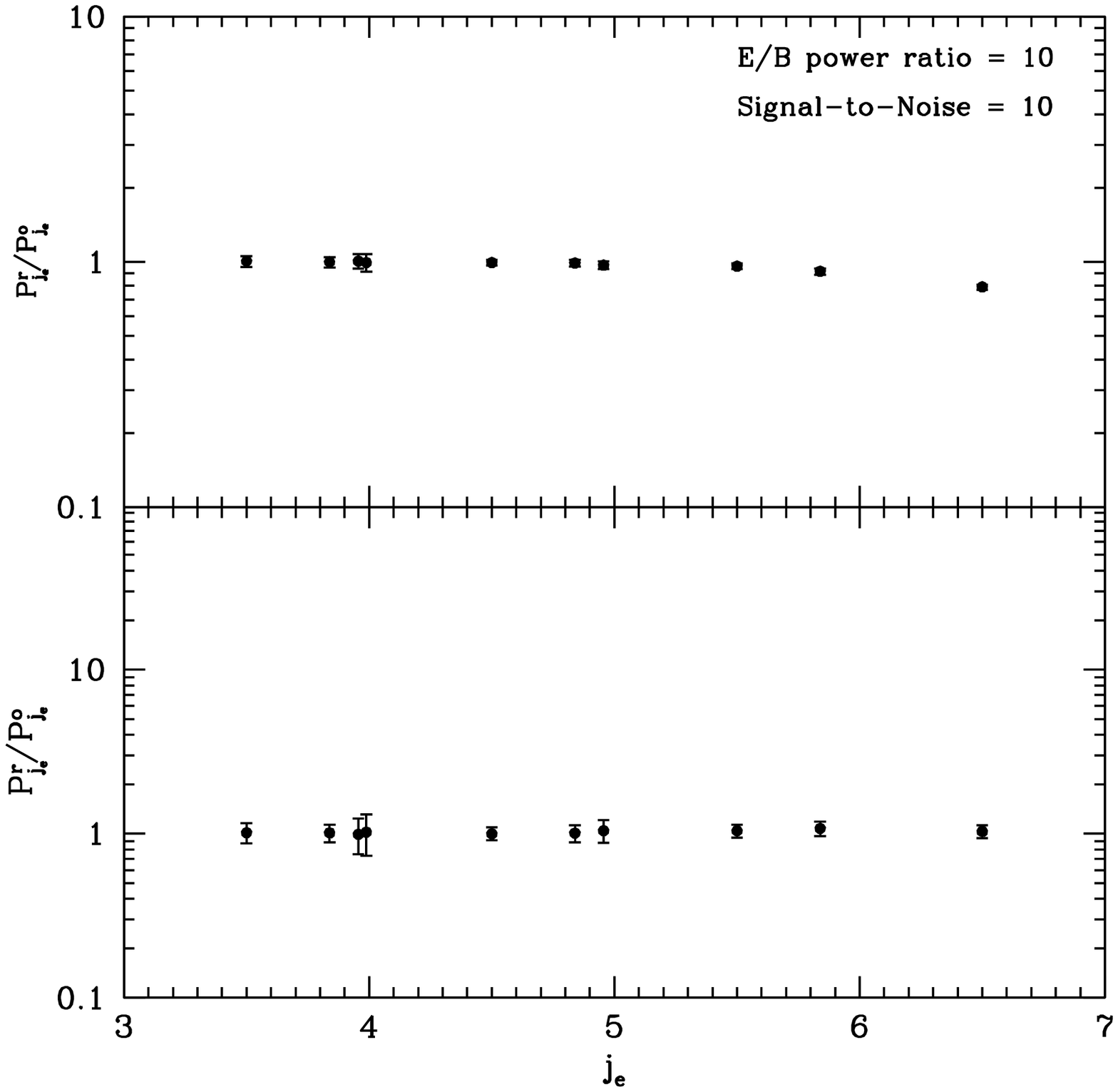}
\end{center}
\caption{The ratio $P^{r}_j/P_j^{o}$ for Gaussian random field
$\psi_{E,B}(x,y)$ with Fourier power spectra $a_{E,B} k^{-3.6}$.
The top panels are for $E$ mode, and bottom for $B$ mode. The
ratio of the powers $E/B$ is equal to 10. The maps $Q$ and $U$ are
added noises with the level of S/N equal to 100 (left), 20
(middle) and 10 (right). }
\end{figure}
\begin{figure}[htb]
\begin{center}
\includegraphics[width=5.0cm]{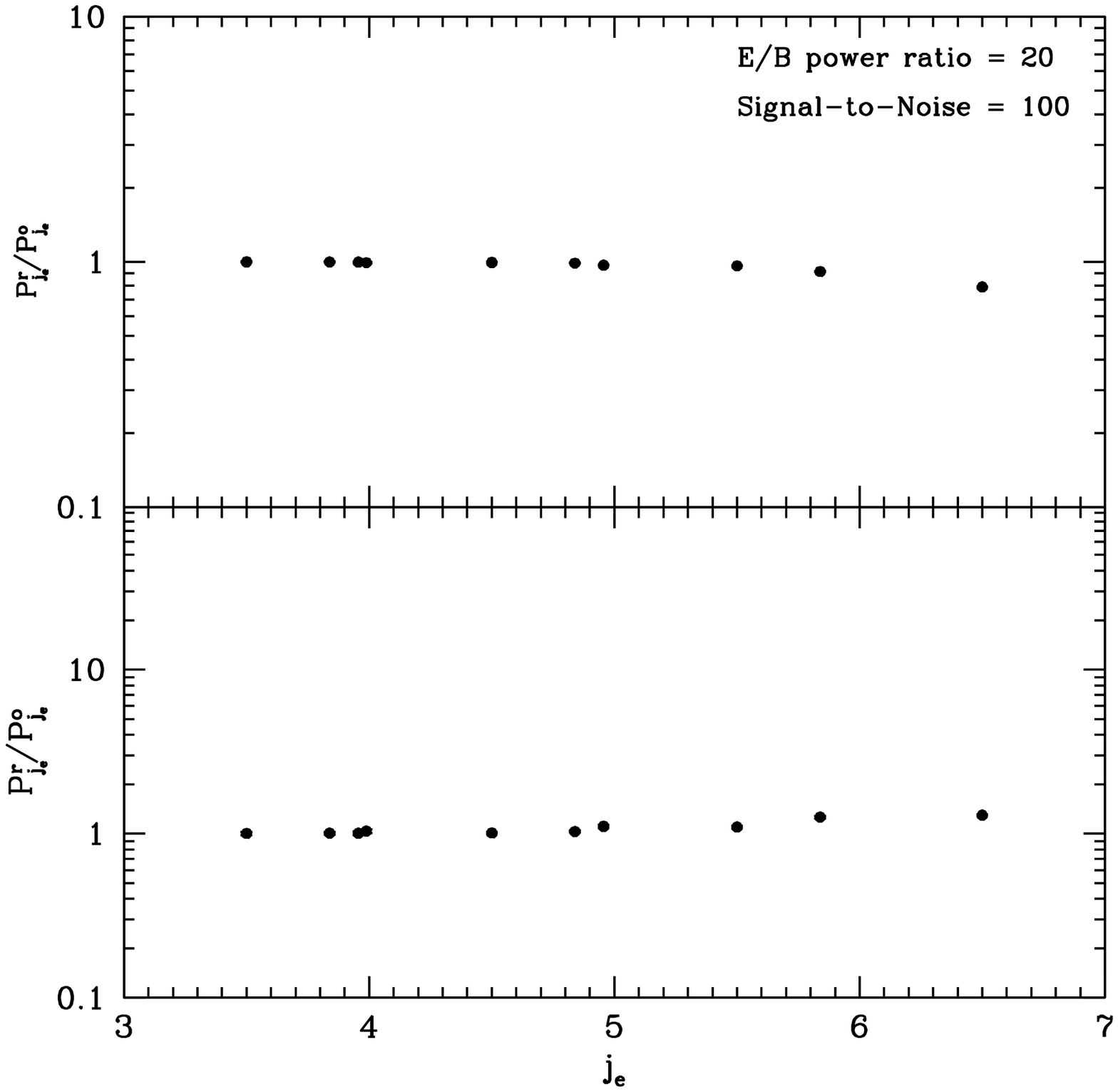}
\hspace{-0.0cm}\includegraphics[width=5.0cm]{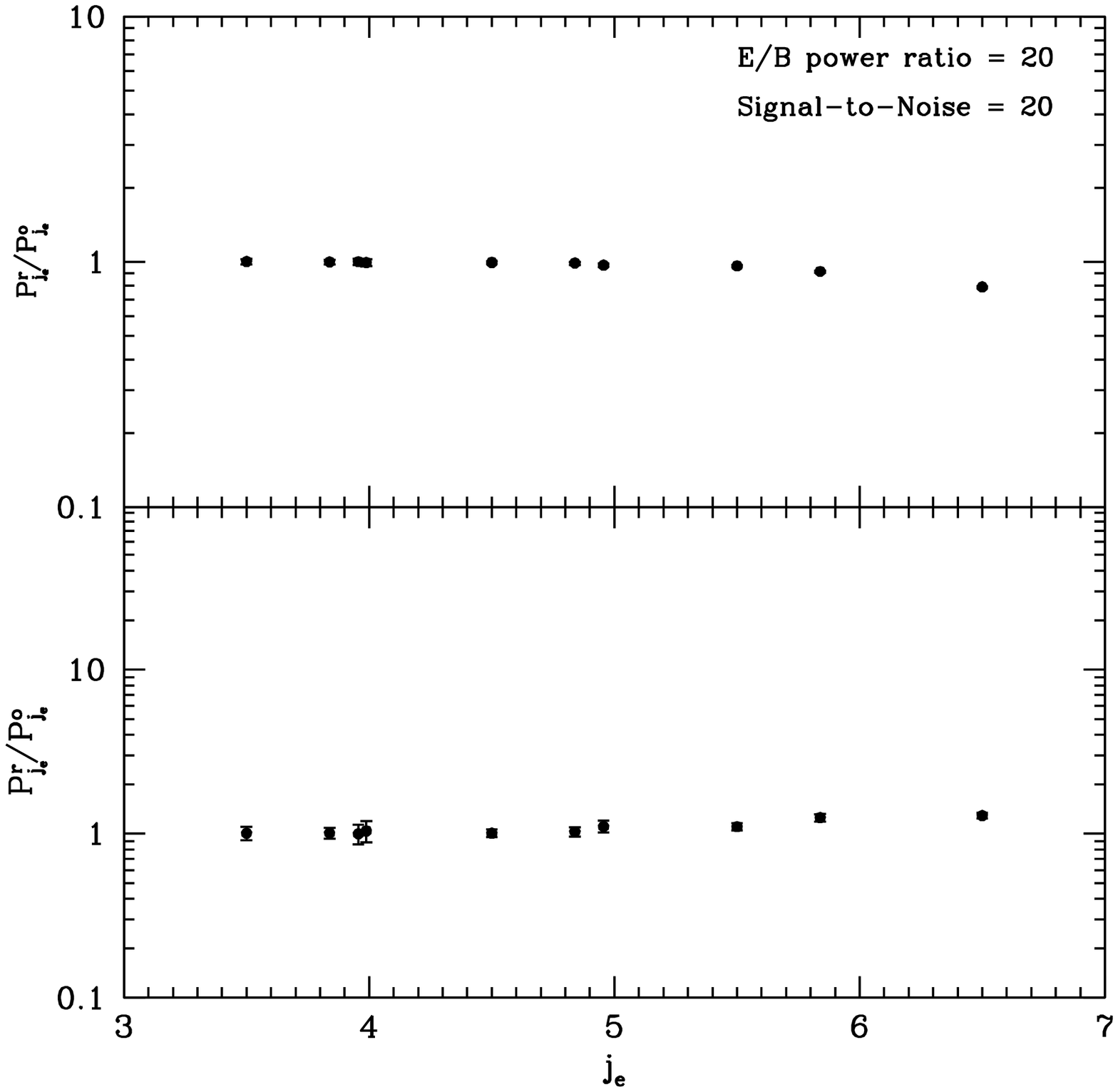}
\hspace{-0.0cm}\includegraphics[width=5.0cm]{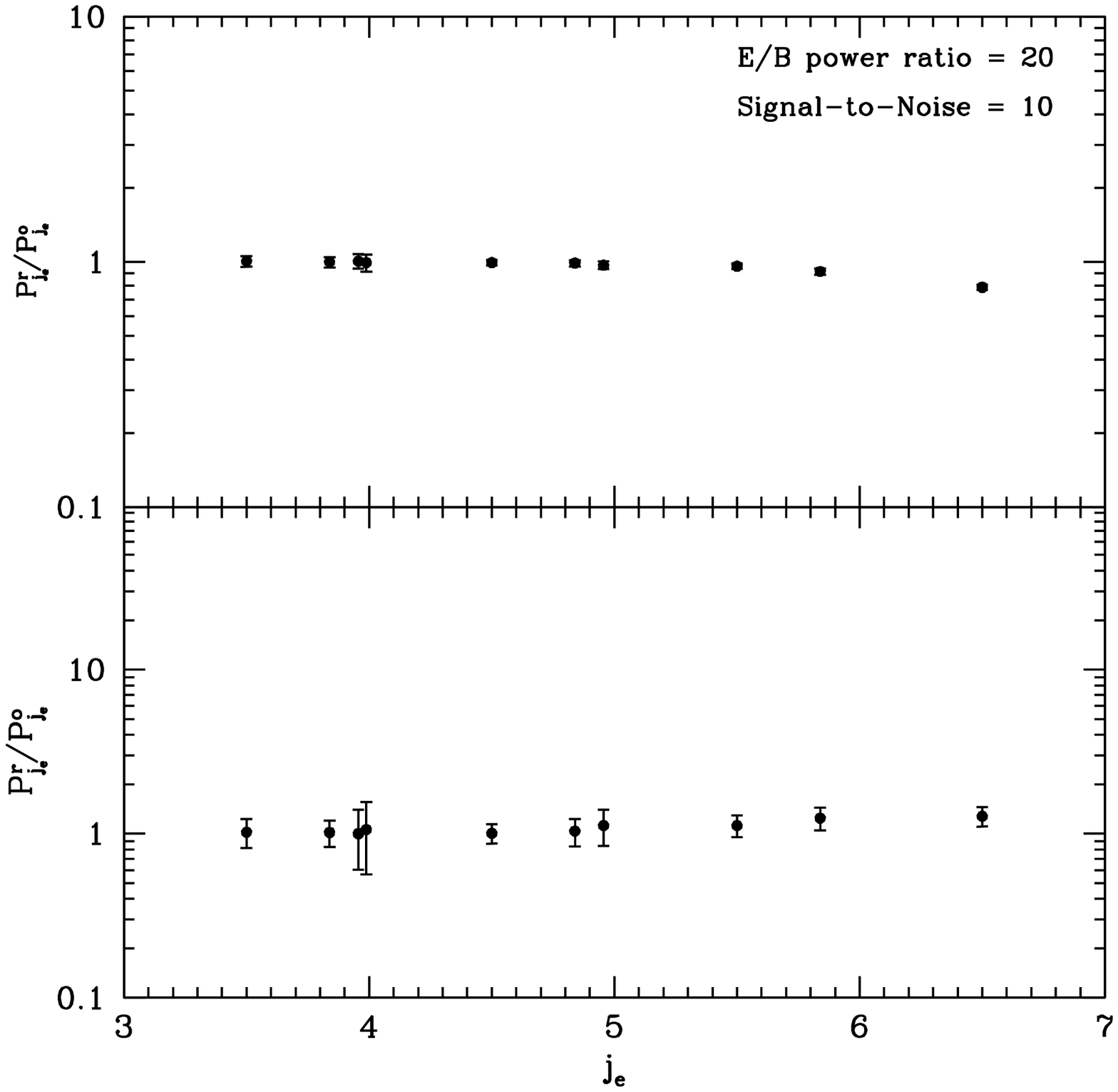}
\end{center}
\begin{center}
\caption{The same as Figure 11, but the power ratio
 of $E$ and $B$ modes is equal to 20.}
\end{center}
\end{figure}
\begin{figure}[htb]
\begin{center}
\includegraphics[width=5.0cm]{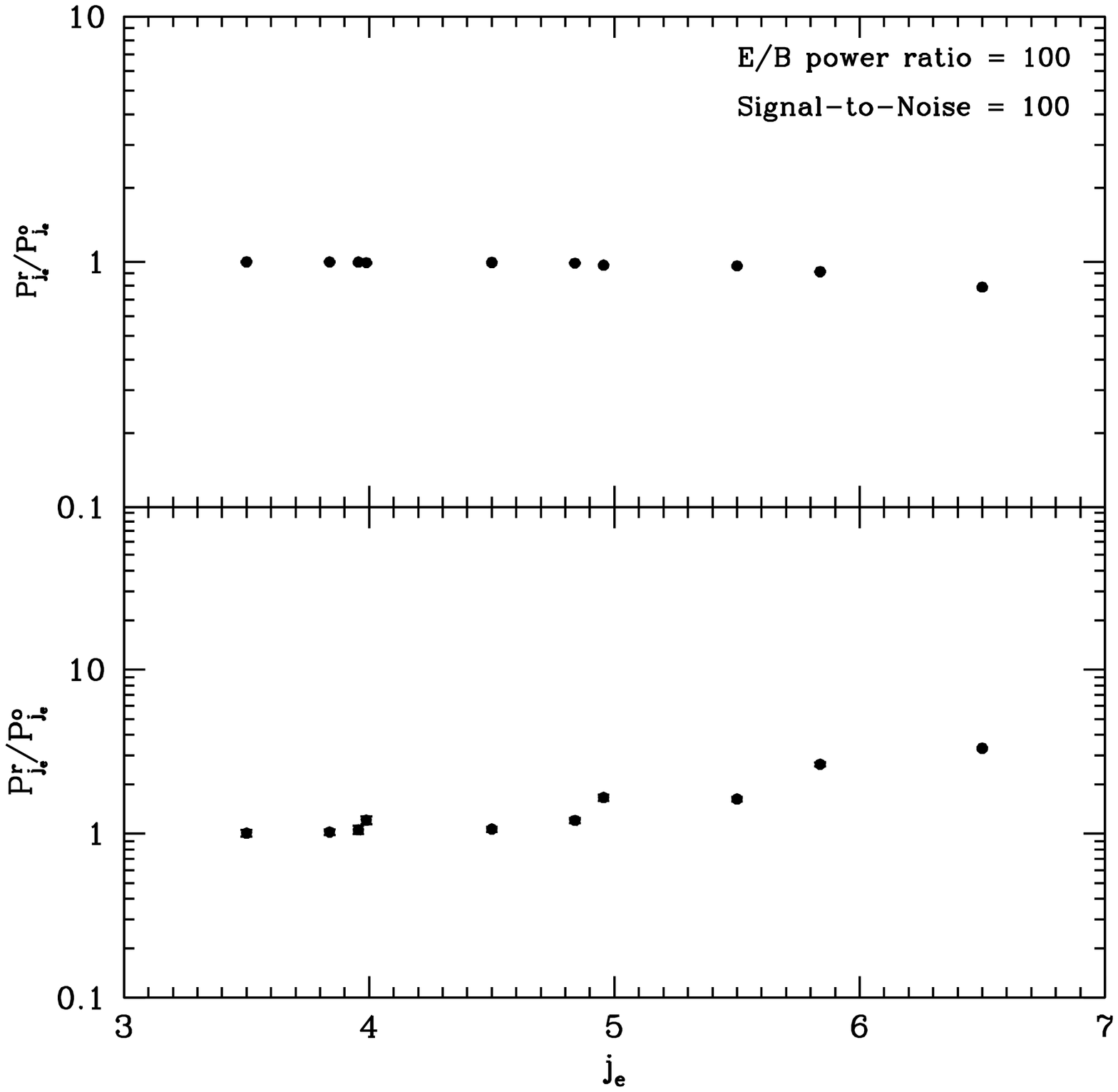}
\hspace{-0.0cm}\includegraphics[width=5.0cm]{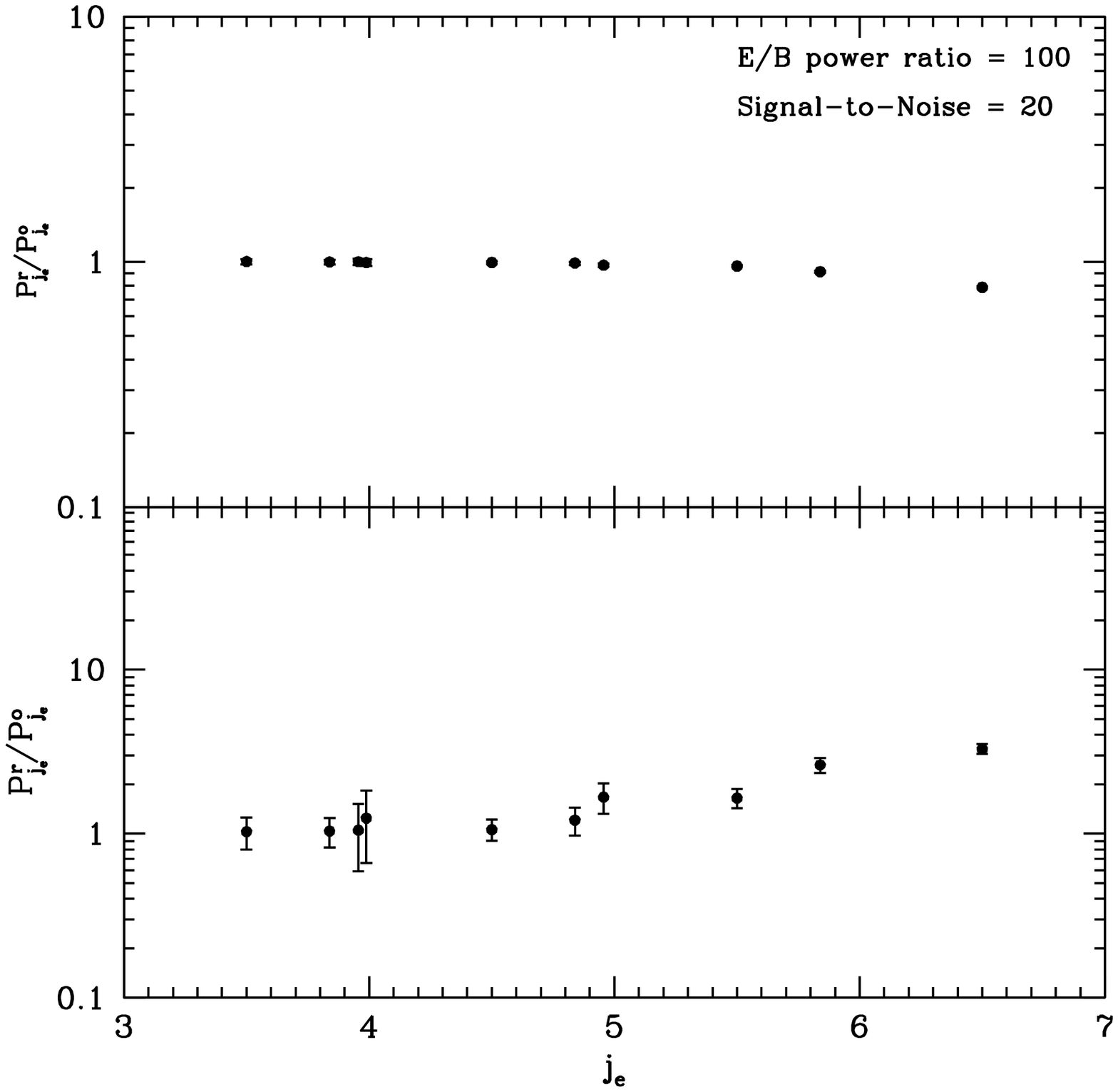}
\hspace{-0.0cm}\includegraphics[width=5.0cm]{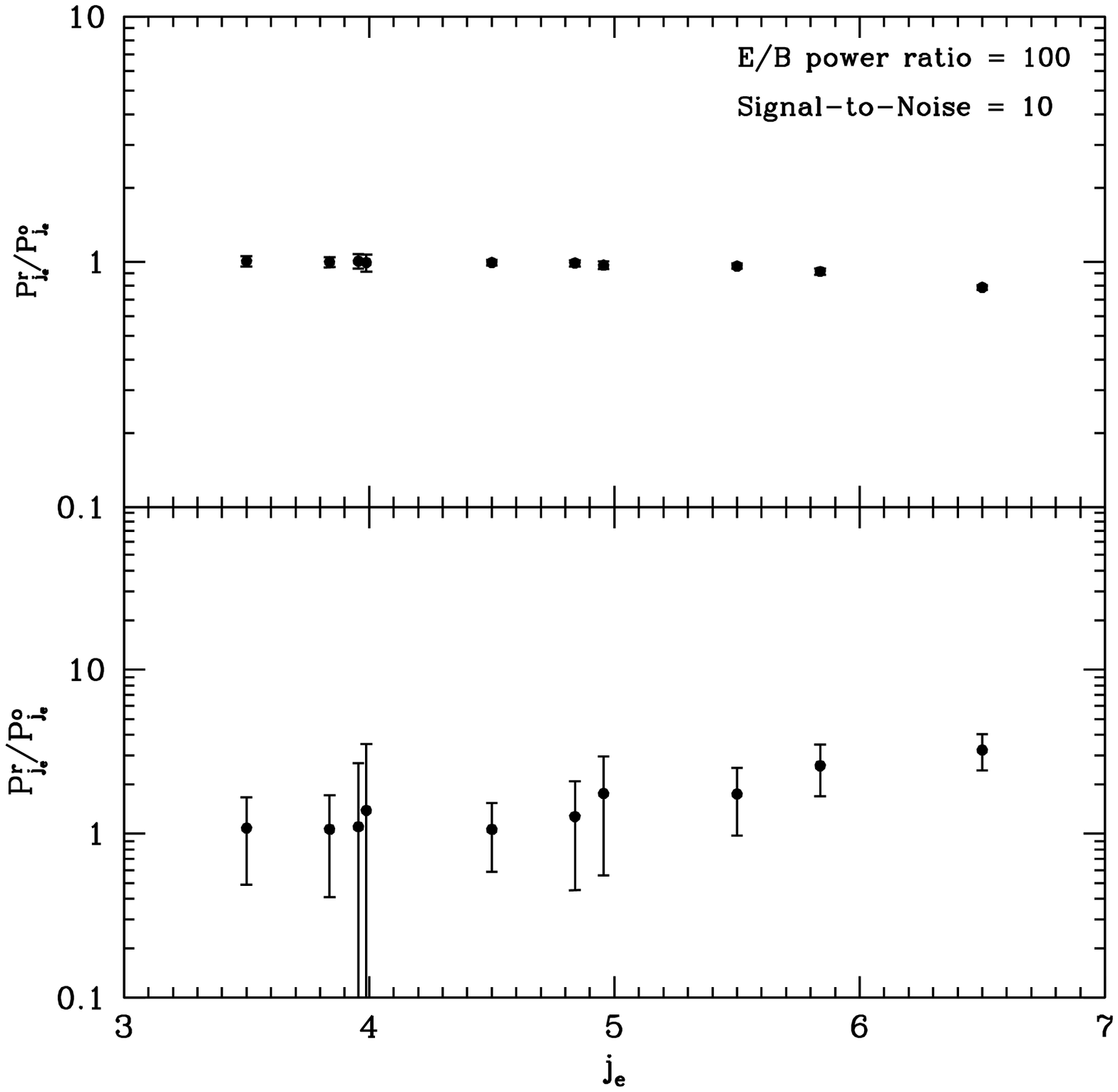}
\end{center}
\begin{center}
\caption{The same as Figure 11, but the power ratio of $E$ and $B$
modes is equal to 100.}
\end{center}
\end{figure}

Figure 11 shows the powers of $E$ mode can be recovered on all
scales $j_{\rm eff}\leq 6$ on all noise levels. On the smallest
scale, $j_{\rm eff}= 6.5$, the ratio $P^{r}_j/P_j^{o}$ of $E$ mode is
slightly lower than 1. This deviation is almost independent of the
level of S/N. Therefore, the errors mostly are not due to the
Gaussian noise addition, but from the effect of leakage. This point
is consistent with the leaking shown in Figure 9, which also give a
little small power on the smallest scale. More interesting, Figure
11 shows that the $B$ mode can be perfectly recovered on all scales
and all noise levels considered.

Figure 12 presents the case of $E/B=20$. The results on scales
$j_{\rm eff}\leq 5.5$ are about the same as the case of $E/B=10$, while
the deviations of $P^{r}_j$ from $P_j^{o}$ on scales $j_{\rm eff}>
5.5$ are larger than that of $E/B=10$. On small scales, the
recovered $E$ powers, $P^{r}_j$, are little lower than the original
power $P_j^{o}$, while the recovered $B$ powers are little higher
than the original power.

Figure 13 is for the case of $E/B=100$. It shows that the recovered
$E$ mode power spectrum is still good on scales $j_{\rm eff}\leq
5.5$ for all S/N. However, on scales $j_{\rm eff} > 5$, the
recovered $B$ mode power spectrum generally is higher than the
original one. This deviation is expected, as Figure 9 shows that the
leaked power from $E$ mode to $B$ mode can be as high as 1\% on
small scales. Nevertheless, the recovered $B$ mode power spectrum is
reasonable on scales $j_{\rm eff} < 5$, even when the S/N is equal
to 10. That is, one can pick up the weak signal of $B$ mode with the
DWT algorithm even when the Gaussian noise level of $Q$ and $U$ maps
is comparable with the $B$ mode signal.

\section{Discussions}

The relationships between the polarization maps of ($E$, $B$) and
($Q$, $U$) are differential. For pixelized samples of $Q$ and $U$ in
finite area, the algorithm of $E/B$ decomposition should be able to
properly handle the derivative operation on a spatially discrete and
noisy data set. The derivative operator $\partial_x$ is a continuous
linear operator to mapping functions defined in Hilbert space, while
the functions $Q$ and $U$ are defined in space spanned by bases
$u_i$, in which $i$ is a set of finite index. Therefore, we should
approximate the derivative operator from mapping between functions
defined in Hilbert space, to a mapping in subspace spanned  by
$n_i$.

What we need to calculate is
\begin{equation}
Of=g,
\end{equation}
where $O$ is a linear continuous operator, like Laplace or
derivative, and $f$ and $g$ are function of $x,y$ in continuous
space $0<x,y<L$. However, we don't know $f$, but only the discretized $\tilde{f}$,
which is given in $N$ pixels (cells). That is, $\tilde{f}$ can be
expressed as
\begin{equation}
\tilde{f}(x,y)=\sum_{k=1}^{N}\alpha_k w_k(x,y),
\end{equation}
where function $w_i(x,y)$ is the binning function of pixel $k$, and
$\alpha_k$ is the observed $f$ at pixel $k$. The simplest binning
function would be the top-hat window function of pixel $k$.

With $N$-dimensional space spanned by bases
$[v_1(x,y),...v_N(x,y)]$, eqs.(36) and (37) yield
\begin{equation}
g_i\equiv \langle g,v_i\rangle=\sum_{k=1}^{N}\langle Ow_k,
v_i\rangle \alpha_k.
\end{equation}
The matrix $O_{ik}\equiv \langle Ow_k, v_i\rangle$ gives a
discretization of operator $O$ from the space $0<x,y<L$ to a
finite-dimensional subspace $[v_i]$.

A Galerkin discretization requires the following equation to be hold
for all $v_i$
\begin{equation}
\langle (g - Of),v_i\rangle=0
\end{equation}
That is, eq.(36) should be hold in the subspace spanned by bases
$[v_1(x,y),...v_N(x,y)]$. In this case, the matrix $O_{ik}$ is a
linear operator to map functions defined in the subspace $v_i$. If
$f$ is a function of the subspace $v_i$, $g=Of$ is also a function
of the subspace.

It can be seen from eqs.(36) - (38) that the discretization of
operator $O$ actually is inevitable for all algorithms. To treat the
data eq.(37), we must use some base $v_i$ in the spatial domain. It
will yield a matrix $O_{ik}$, regardless whether eq.(39) is hold
with the bases $v_i$. The Galerkin method gives a best
discretization of eq.(36) or the operator $O$ (e.g. Louis et al.
1997).

We use $\phi_{J,l}(x)\phi_{J,l'}(y)$, $l,l'=0...2^J-1$ to be the
bases to span the subspace with dimension $N=2^{J}\times 2^J$.  It
can be shown that the conditions of Galerkin discretization,
eq.(39), will be satisfied for operator $O=\partial_x$,
$\partial_y$, $\partial^2_x$ and $\partial^2_y$. That is, for any
function $f(x,y)$ of the subspace spanned by bases
$\phi_{J,l}(x)\phi_{J,l'}(y)$, $l,l'=0...2^J-1$, the result of $Of$
are also functions of the subspace. This is the wavelet-Galerkin
discretization. Matrix $\langle Ow_k, v_i\rangle$ will be invertible.
In this sense, the discretization does not lose information, or
introduces false data or correlations.

Obviously, the Galerkin discretization is not unique. One can use
different wavelets to do the Galerkin discretization. To apply the
discretization, the matrix $O_{ik}\equiv \langle Ow_k, v_i\rangle$
should have the following desirable properties. First, the matrix
$O_{ik}$ has to be sparse, narrowly banded. In this case, one can
effectively minimize the information lose due to dropping boundary
modes. A narrowly banded matrix can also effectively reduce the
spreading of errors among cells with different ${\bf l}$. Secondly,
in order that the errors not increase with the size of the matrix
$N$ (number of data), the ``width" of the band in which the matrix
elements $O_{ik}$ is non-zero, should be independent on $N$.

\section{Conclusions}

The algorithm developed in this paper can be summarized as follows
\begin{itemize}
\item From observed noisy and discrete maps $Q(x,y)$ and $U(x,y)$ we calculate
their DWT maps $Q_{l_1,l_2}$ and $U_{l_1,l_2}$ on the finest scale
${\bf j}=(J,J)$, which is given by the resolution.
\item Using eqs.(16) and (17) we decompose $Q_{l_1,l_2}$ and $U_{l_1,l_2}$
into $\mathbb{E}_{l_1,l_2}$ and $\mathbb{B}_{l_1,l_2}$.
\item Using eqs.(19) and (20) we calculate WFCs $\tilde{\epsilon}^E_{\bf j,l}$
and $\tilde{\epsilon}^B_{\bf j,l}$ on scales $(j_1,j_2)$ and
$j_1,j_2 \leq J$.
\item Using the WFC maps $\tilde{\epsilon}^E_{\bf j,l}$  and
$\tilde{\epsilon}^B_{\bf j,l}$ we calculate the DWT power spectra by
eqs.(30) and (31).
\item We identify spatial structures with maps of $\mathbb{E}_{l_1,l_2}$
and $\mathbb{B}_{l_1,l_2}$.
\end{itemize}

With this algorithm, it is possible to recover the power spectrum of
$B$-mode random fields from noisy Stokes parameter maps $Q$ and $U$
when the power ratio $E/B$ is as high as $10^2$, and the S/N is
equal to or higher than 10. For samples with given structures, the
$B$-mode structure can also be identified when the power ratio $E/B$
is equal to 10$^2$. Besides power spectrum, the DWT variables of
SFCs ($\epsilon^E_{\bf j,l}$, $\epsilon^B_{\bf j,l}$) and WFCs
($\tilde{\epsilon}^E_{\bf j,l}$, $\tilde{\epsilon}^B_{\bf j,l}$) can
be used for high order statistics, such as high order moments,
scale-scale correlation, cross correlation between the $E$ and $B$
and other maps.

With the DWTs, one can construct orthogonal, divergence-free vector
wavelets. It has been used for a local analysis of the velocity
field of incompressible turbulence (Urban 1995; Kishida et al. 1999;
Albukrek et al. 2002). The divergence-free $B$ field is similar to a
2-D velocity field of turbulence of incompressible fluid (e.g. Pina
1998). Therefore, it would be valuable to further study the DWT
$E/B$ decomposition with the divergence-free vector wavelets.

\acknowledgments We thank Dr. Priya Jamkhedkar for her helps.
Liang Cao acknowledge the support by the Knowledge Innovation
Program of the Chinese Academy of Sciences(0990611009) and the CAS
Special Grant for Postgraduate Research, Innovation and
Practice(0992921009). This is work is partially supported by
NSFC(10533030, 10878001), US NSF AST-0507340, and ICRAnet grant
2008.

\appendix

\section{Derivative operator in wavelet representation}

In the DWT space, the operators of derivatives are represented as a
matrix
\begin{equation}
T^{(n)}_{j;l,l'}\delta_{j,j'}=\int
\phi_{j,l}(x)\partial^n_x\phi_{j',l'}(x)dx.
\end{equation}
That is, the matrix is diagonal with respect to $j,j'$.
$T^{(n)}_{j,l,l'}$ is given by (Beylkin 1992; Kwon 1998)
\begin{equation}
T^{(n)}_{j,l,l'}=\frac{1}{h^n}r^{(n)}_{l-l'}
\end{equation}
where $h=1/2^j$. The matrix elements $r^{(n)}_{l-l'}$ depend on the type of
wavelet. For Daubechies 6 wavelet, the non-zero coefficients
$r^{(n)}_{l-l'}$ are $|l-l'|\leq 4$. The values of $r^{(n)}_{l-l'}$
are listed in Table 1, in which $m=l-l'$. Therefore the coefficients
of $T^{(n)}_m$ of eqs.(11) and (12) are given by
\begin{equation}
T^{(n)}_m=\frac{1}{h^n}r^{(n)}_{m}.
\end{equation}

It is interesting to see that the non-zero band of first and second
order derivative operators $\partial_x$ and $\partial^2_x$ are the
same. This is different from the estimation of derivative operator
by differential approximation.

\begin{center}
\begin{tabular}{ccc}
\multicolumn{3}{c}{Table 1 \ Coefficient of $r^{(n)}_{m}$}\\ \hline
 $l-l'=m$ & $n=1$ & $n=2$ \\ \hline
 4  & 1/2920   & 3/560   \\
 3  & 16/1095  & 4/35    \\
 2  & -53/365  & -92/105 \\
 1  & 272/365  & 356/105 \\
 0  & 0        & -295/56 \\
 -1 & -272/365 & 356/105 \\
 -2 & 53/365   & -92/105 \\
 -3 & -16/1095 & 4/35    \\
 -4 & -1/2920  & 3/560   \\ \hline
\end{tabular}
\end{center}


\end{document}